\documentclass[11pt]{article}
\usepackage[dvips]{graphicx}
\def\lsim{\buildrel {\textstyle <}\over {_\sim}}
\begin{document}
\twocolumn[
\baselineskip 4.2mm
\begin{center}
{\bf\Large 
Spatiotemporal Correlations of Earthquakes 
}
\bigskip

{\bf\large Hikaru Kawamura}
\bigskip

{\small\it Department of Earth and Space Science, Faculty of Science,
Osaka University, 
Toyonaka 560-0043, Japan}
\end{center}
%
%
%
%
Statistical properties of earthquakes are studied both 
by the analysis of real earthquake catalog of Japan
and by numerical computer
simulations of the spring-block model in both one and two dimensions.
Particular attention is paid to the spatiotemporal correlations  
of earthquakes, {\it e.g.\/}, the recurrence-time distribution or 
the time evolution before and after 
the mainshock of seismic distribution functions, 
including the magnitude distribution and
the spatial seismic distribution.
Certain eminent features of the spatiotemporal correlations, {\it e.g.\/},
foreshocks, aftershocks, swarms and doughnut-like seismic pattern, 
are discussed.
\bigskip\bigskip
]
%
%
%
%
\baselineskip 4.3mm

\noindent
\section{Introduction}

Although earthquakes are obviously complex phenomena,
the basic physical picture of earthquakes seems to have been well
established now:
Earthquake is a stick-slip frictional instability of a fault
driven by steady motions of tectonic plates \cite{HK:ScholzRev,HK:ScholzBook}.
Although it remains to be extremely difficult at the present stage to say 
something really credible for each
individual earthquake event, if one collects many events and take average over 
these events, a clear tendency often shows up there.
Thus, it is sometimes possible to say something credible for the
{\it average\/} or {\it statistical\/} properties of earthquakes, 
or more precisely, sets of earthquakes. 

In this article, we wish to review some of 
our recent studies on the statistical properties of earthquakes.
Our study of earthquakes
is motivated by the following three issues.

\medskip
\noindent
{\it Critical versus characteristic\/}

It has long been known empirically that certain
power-laws often appear in the statistical properties of earthquakes,
{\it e.g.\/}, the Gutenberg-Richter (GR) law for the magnitude distribution of
earthquakes, or
the Omori law for the time evolution of the frequency of 
aftershocks\cite{HK:ScholzBook}. Power-law means that there is no
characteristic scale in the underlying physical
phenomenon. In statistical physics, 
one of the most widely recognized occasion of the appearance of a power-law 
is critical phenomena associated with a thermodynamic
second-order phase transition. 
Indeed, inspired by this analogy, Bak and collaborators
introduced the concept of the ``self-organized criticality (SOC)'' into 
earthquakes\cite{HK:Bak,HK:BakTang}. 
According to this picture, The Earth's crust is always in the 
critical state generated dynamically, and power-laws associated 
with statistical properties of earthquakes are regarded as 
realizations of the 
intrinsic critical nature of earthquakes. Indeed,
the SOC idea gives a natural
explanation of the scale-invariant power-law behaviors frequently observed in
earthquakes, including the GR law and the Omori law. 

In contrast to such an SOC view of earthquakes, 
an opposite view has also been common in earthquake studies,
a view which regards earthquakes  as ``characteristic''. 
In this view, earthquakes are supposed to possess their own 
characteristic scales, {\it e.g.\/}, a characteristic energy scale
or a characteristic time scale.

Thus, whether an earthquake is critical or characteristic remains to be
one of central issues of modern earthquake studies.

\medskip
\noindent
{\it Possible precursory phenomena --- spatiotemporal correlations 
of earthquakes\/}

In conjunction with earthquake prediction, possible precursory
phenomena associated with large earthquakes have special importance.
If one takes a statistical approach, a natural quantity to be examined
might be spatiotemporal correlations of seismicity, {\it i.e.\/}, 
how earthquakes correlate in space and time. 
If we could identify the property in which
a clear anomaly is observed preceding the large event, it might be 
useful for
earthquake prediction. In the present article, we wish to
investigate among others various types of spatiotemporal 
correlation functions of earthquakes.

\medskip
\noindent
{\it The constitutive relation and the nature of stick-slip dynamics}

Since earthquakes can be regarded as a stick-slip frictional 
instability of
a pre-existing fault, the statistical properties of earthquakes 
should be governed  by the physical law of 
rock friction\cite{HK:ScholzRev,HK:ScholzBook}. Unfortunately,
our present understanding of physics of friction is still poor.
We do not have precise knowledge of the constitutive relation governing
the stick-slip dynamics at earthquake faults.
The difficulty  lies partly in 
the fact that a complete microscopic theory of 
friction is still not available, but also in the fact that the length and
time scales relevant to earthquakes are so large that the applicability
of laboratory experiments of rock friction is not necessarily clear.

A  question of fundamental importance in earthquake studies
might be how the properties of earthquakes depend on the constitutive
relation and the material parameters 
characterizing earthquake faults, 
including the elastic properties of the crust and the constitutive
parameters characterizing the friction force.

To answer this question and to get deeper insight into the physical 
mechanism governing the stick-slip process of earthquakes, 
a proper modeling of an earthquake
might be an important step. In fact,
earthquake models of various
levels of simplifications  have been proposed in geophysics
and statistical physics, and their statical properties have been extensively
studied mainly by means of numerical computer simulations.
One of the standard model is the so-called 
spring-block model originally proposed by Burridge and Knopoff 
(BK model) \cite{HK:BK}, which
we will employ in the present particle.
Yet, our present understanding of the question  
how the earthquake properties depend on the constitutive relation
and the material
parameters characterizing earthquake faults remains far from satisfactory.

\medskip
In the present article,
in order to approach the three goals mentioned above,
we take two complementary approaches:  
In one, we perform numerical computer simulations based on the 
spring-block model
to clarify the spatiotemporal correlations of seismic events.
Both the one-dimensional (1D) BK model \cite{HK:MoriKawamura,HK:MoriKawamura2}
and the two-dimensional (2D) BK model \cite{HK:MoriKawamura3} are studied.
In the other, we analyze the earthquake catalog of Japan
to examine the spatiotemporal correlations of real seismicity
\cite{HK:KawamuraMori}.
We then compare the results of numerical model simulation and 
the analysis of real  earthquake catalog,
hoping that such a comparison   might give us useful information
about the nature of earthquakes.

The following part of the article is organized as follows.
In section 2, we introduce
the model employed in our numerical computer simulations, and
explain some the details of the simulations. We also introduce
the earthquake catalog used in our analysis of real seismicity
of Japan. Then, in section 3, 
we report on the results of our analysis of the
spatiotemporal correlations of real earthquakes  
based on the earthquake catalog of Japan, together
with the results of our numerical simulations of the 1D and 2D
BK models. The statistical properties examined in this section include; 
(i) the magnitude distribution,
(ii) the local recurrence-time distribution, (iii) the global recurrence-time 
distribution, (iv) the time evolution of the spatial seismic distribution 
before the mainshock,
(v) the time evolution of the spatial seismic distribution 
after the mainshock, and (vi) the 
time-resolved magnitude distribution before and after the mainshock.
Finally, section 4 is devoted to summary and discussion of our results.

\section{The model, the simulation 
and the catalog}

In this section, we introduce the spring-block model which we will
use in our numerical computer simulation, and explain 
some of the details of the simulation. We also introduce the
seismic catalog of Japan which we will use in our analysis of the 
spatiotemporal
correlations of real earthquakes. The results of our numerical computer
simulations and the analysis of seismic catalog of Japan
will be presented in the following section 3.

\subsection{The spring-block model of earthquakes}

The spring-block model of earthquakes was
originally proposed by Burridge and Knopoff 
\cite{HK:BK}.
In this model, an earthquake fault is
simulated by an assembly of blocks, each of which is connected via 
the elastic springs to the neighboring blocks and to 
the moving plate. 
All blocks are subject to the 
friction force, the source of the nonlinearity in the
model, which eventually realizes an earthquake-like frictional instability. 
The model contains several parameters representing, {\it e.g.\/},  
the elastic properties of the crust and the
frictional properties of faults.

In the 1D BK model, the equation of motion for the $i$-th block can be
written as
\begin{equation}
m \ddot U_i=k_p (\nu ' t'-U_i) + k_c (U_{i+1}-2U_i+U_{i-1})-\Phi_i,
\label{HK:Eq1}
\end{equation}
where $t'$ is the time, $U_i$ is the displacement of the 
$i$-th block, 
$\nu '$ is the loading rate 
representing the speed of the plate, and $\Phi_i$ is the friction force
working at the block $i$. 

In order to make the equation
dimensionless, we measure the time $t'$ in units of the characteristic 
frequency $\omega =\sqrt{k_p/m}$ and the displacement $U_i$ in units of
the length $L=\Phi_0/k_p$, $\Phi_0$ being the static friction. Then,
the equation of motion can be written  in the dimensionless form as
\begin{equation}
\ddot u_i=\nu t-u_i+l^2(u_{i+1}-2u_i+u_{i-1})-\phi_i,
\label{HK:Eq2}
\end{equation}
where $t=t'\omega $ is the dimensionless time, 
$u_i\equiv U_i/L$ is the dimensionless displacement of the 
$i$-th block, 
$l \equiv \sqrt{k_c/k_p}$ is the dimensionless stiffness parameter, 
$\nu =\nu '/(L\omega)$ is the dimensionless loading rate, and  
$\phi_i \equiv \Phi_i/\Phi_0$ is the dimensionless friction 
force working at the block $i$. 

The form of the friction force $\phi$ is specified by the constitutive 
relation. As mentioned, this part is a vitally important, yet largely
ambiguous part in the proper description of earthquakes.
In order for the model to exhibit a 
dynamical instability corresponding to an earthquake, it is essential 
that the friction force $\phi$ possesses a frictional {\it weakening\/}
property, {\it i.e.\/},
the friction should become weaker as the block slides.

As the simplest form of the friction force, 
we assume here 
the form used by Carlson {\it et al\/}, which represents the
velocity-weakening friction force \cite{HK:CLST,HK:CLSRev}. 
Namely, the friction force $\phi(\dot u)$ 
is assumed to be a single-valued function of the velocity $\dot u_i$,
{\it i.e.\/}, $\phi_i$,
gets smaller as $\dot u_i$ increases,
\begin{equation}
\phi(\dot u) = \left\{ 
             \begin{array}{ll} 
             (-\infty, 1],  & \ \ \ \ {\rm for}\ \  \dot u_i\leq 0, \\ 
              \frac{1-\sigma}{1+2\alpha \dot u_i/(1-\sigma )}, &
             \ \ \ \ {\rm for}\ \  \dot u_i>0, 
             \end{array}
\right.
\label{HK:Eq3}
\end{equation}
where its maximum value corresponding to the static friction
has been normalized to unity. As noted above, this normalization 
condition $\phi(\dot u=0)=1$ has been utilized to set the length unit $L$.
The back-slip is inhibited by imposing an
infinitely large friction for $\dot u_i<0$, {\it i.e.\/}, 
$\phi(\dot u<0)=-\infty $. 

In this velocity-weakening constitutive relation, 
the friction force is characterized by the two parameters, $\sigma$ and 
$\alpha$. The former, $\sigma$, 
represents an instantaneous drop of the friction force
at the onset of the slip, while the latter, $\alpha$, 
represents the rate of the friction force getting weaker
on increasing the sliding velocity. In our simulation, 
we regard $\sigma$ to be small, and fix $\sigma =0.01$. 

It should be emphasized again that, although the above Carlson-Langer
velocity-weakening friction is rather simple and has been widely used 
in numerical simulations, the real 
constitutive relation might not be so simple with features
possibly different from the simplest velocity-weakening one.
Indeed, there have been several other proposals for the law of rock
friction, {\it e.g.\/}, the slip-weakening friction 
force \cite{HK:ScholzRev,HK:ScholzBook,HK:Shaw95,HK:Myers96} 
or the rate- and state-dependent 
friction force \cite{HK:ScholzRev,HK:ScholzBook,HK:Dietrich,HK:Ruina,HK:TseRice,HK:Kato,HK:OhmuraKawamura}. 
In this article,  we leave the study of these different constitutive 
relations to other references 
or to future studies, and assume the simplest velocity-weakening 
friction force given above.
 
We also assume the loading rate $\nu$ to be infinitesimally small, and put 
$\nu=0$ during an earthquake event, a very good approximation 
for real faults \cite{HK:CLST,HK:CLSRev}. Taking this limit ensures that 
the interval time during successive
earthquake events can be measured in units of $\nu^{-1}$ 
irrespective of 
particular values of $\nu$. Taking the $\nu \rightarrow 0$ limit
also ensures that, during an ongoing event,
no other event takes place at a distant 
place, independently of this ongoing event. 

The extension of the 1D BK model to 2D is rather straightforward.
In 2D, the blocks are considered to be arranged in the form of a
square array connected with the springs of the spring constant
$k_c$ \cite{HK:Carlson91b}. We consider the isotropic and uniform 
case where $k_c$ is uniform everywhere in the array
independent of the spatial directions.
All blocks are connected to 
the moving plate via the springs of the spring constant $k_p$, and are also
subject to the velocity-weakening friction force defined above. 
The plate is driven
along the $x$-direction with a constant rate $\nu$. It is assumed
that all blocks move along the $x$-direction only, {\it i.e.\/}, 
the displacement of the block at the position $(i_x,i_y)$ is assumed to
be given by
$\vec u(i_x, i_y)=(u_x(i_x,i_y), 0)$.

Since the early study by Burridge and Knopoff \cite{HK:BK}, the properties of
this BK model has been studied.simulations,

Carlson, Langer and collaborators
performed a pioneering study of 
the statistical properties of 
the 1D BK model quite extensively
\cite{HK:CLST,HK:CLSRev,HK:Carlson91b,HK:CL,HK:Carlson91a}, 
with particular attention to
the magnitude distribution of earthquake events.
It was observed that,
while smaller events persistently obeyed the GR law, {\it i.e.\/},
staying critical or near-critical, 
larger events exhibited a significant 
deviation from the GR law, being off-critical 
or ``characteristic'' \cite{HK:CLST,HK:Carlson91b,HK:CL,HK:Carlson91a,HK:Schmittbuhl}.  
Shaw, Carlson and Langer studied the same model by
examining the spatiotemporal patterns
of seismic events preceding large events, observing that
the seismic activity accelerates as the large event approaches
\cite{HK:Shaw92}.

The BK model was also extended in several ways, 
{\it e.g.\/}, taking account of the effect of the viscosity 
\cite{HK:Myers93,HK:Shaw94,HK:Anantha},  modifying the form of the friction 
force
\cite{HK:Myers93,HK:Anantha}, or taking account of the long-range 
interactions \cite{HK:Rundle}. 
The 2D version of the BK model was also analyzed by Carlson 
\cite{HK:Carlson91b} and by Myers {\it et al\/} \cite{HK:Myers96}.

The study of statistical properties of earthquakes was promoted in early 
nineties, inspired by the  work by P. Bak and collaborators
who emphasized the concept of 
``self-organized criticality (SOC)'' \cite{HK:Bak,HK:BakTang}. 
The SOC idea was developed mainly on the basis of  
the cellular-automaton versions of the earthquake 
model \cite{HK:Bak,HK:BakTang,HK:Nakanishi,HK:Ito,HK:Brown,HK:OFC,HK:Hergarten,HK:Hainzl,HK:Helmstetter}.
The statistical properties of these cellular-automaton models
were also studied quite extensively, often
interpreted within the SOC framework. These models apparently reproduce
several fundamental features of earthquakes such as the GR law, the 
Omori law of aftershocks, the existence
of foreshocks, {\it etc\/}. We note that,
although many of these cellular-automaton models were  originally
introduced to mimic the spring-block model, 
their statistical properties
are not always identical with the original spring-block model.
Furthermore, as compared with the spring-block model,
the cellular-automaton models are much more simplified so that
the model does not have  enough room to represent various material 
properties of the earthquake fault in a physically appealing way.
Thus,  the spring-block model has an advantage over the cellular-automaton
models 
that the dependence on the
material parameters, including  the constitutive and elastic properties, 
are more explicit.

\subsection{The numerical simulation}

As already mentioned, numerical model simulation is a quite useful tool
for our purposes:
First, generating huge number of events required to attain the 
high precision for discussing the statistical properties of
earthquakes is
easy to achieve in numerical simulations whereas it
is often difficult to achieve in real earthquakes, especially
for larger ones.
Second,  various material parameters characterizing earthquake faults
are extremely 
difficult to control in real faults,  whereas they are easy to control
in model simulations.

We solve the equation of motion (2) 
by using the Runge-Kutta method of the fourth 
order. The width of the time discretization $\Delta t$ is taken to be
$\Delta t\nu =10^{-6}$. We have checked that the statistical properties 
given below are unchanged
even if we take the smaller $\Delta t$. Total number of $10^7$  events
are generated in each run, 
which are used to perform various averagings. In calculating the observables,
initial $10^4$ events are 
discarded as transients.

In order to eliminate the possible finite-size effects, the total
number of blocks $N$ are taken to be large. 
In 1D, we set $N=800$, imposing the
periodic boundary condition. The size dependence of the results was
examined in Ref.\cite{HK:MoriKawamura2} with varying $N$ in the range
$800\leq N \leq 6400$.
In 2D, we set our lattice size $160 \times 80$ 
with the periodic boundary condition on the longer side, and the free boundary
condition on the shorter side.

We study the properties of the model, with varying the frictional parameter
$\alpha$ and the elastic parameter $l$. In this article,
attention is paid to
the dependence on the parameter 
$\alpha$, since the parameter $\alpha$, which represents the
extent of the frictional weakening,  
turns out to affect the result most significantly \cite{HK:MoriKawamura2}.

\subsection{The seismic catalog of Japan}

\begin{figure}[ht]
\begin{center}
\includegraphics[scale=0.45]{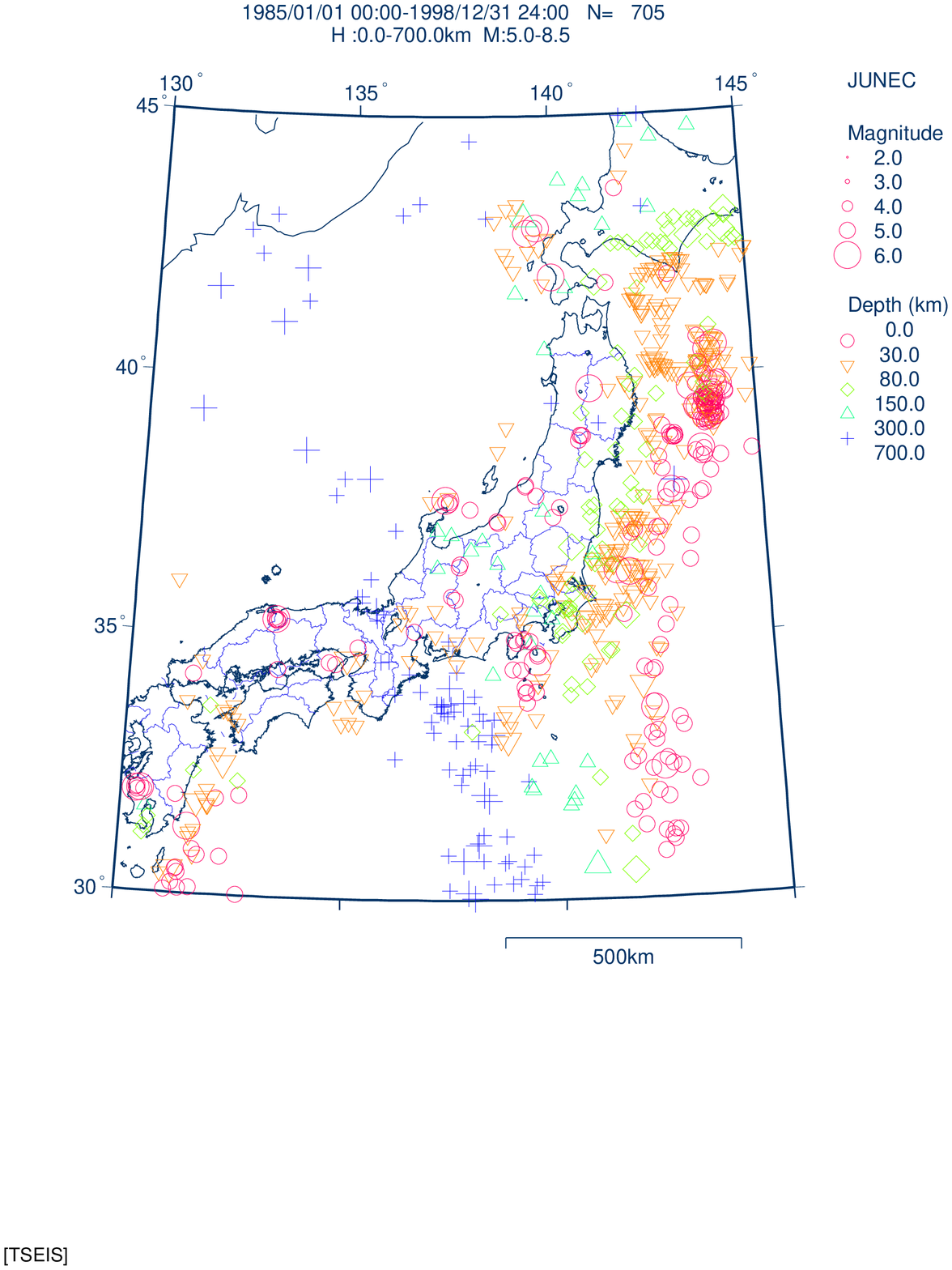}
\end{center}
\vspace{-3cm}
\baselineskip 3.2mm
\caption{\small 
A seismicity map of Japan generated from the JUNEC catalog. Large earthquakes 
of their magnitudes greater than five, which
occurred in Japan area during 1985-1998, are mapped out by using the program
developed by H. Tsuruoka \cite{HK:Tsuruoka}.
}
\label{HK:fig1}
\end{figure}

In order to compare the simulation data on the 1D and 2D BK 
models with the corresponding data for real earthquakes,
we analyze in the following section the seismic catalog of Japan provided by
Japan University Network Earthquake Catalog (JUNEC) available at
http://kea.eri.u-tokyo.ac.jp/CATALOG/junec/monthly.html.
The catalog covers earthquakes which occurred in Japan area
during July 1985 and December 1998. 
Total of 199,446 events are recorded in the catalog. As an example,
we show in Fig.\ref{HK:fig1}, 
a seismicity map of Japan generated from the JUNEC catalog, 
where all large earthquakes 
of their magnitudes greater than five, which
occurred in Japan area during 1985-1998, are mapped out \cite{HK:Tsuruoka}.

\section{Statistical properties of earthquakes}

In this section, we show the results of our numerical simulations of the
1D and 2D BK models, and compare them 
with the results of our analysis of the seismic
catalog of Japan (JUNEC catalog). We study
several observables, {\it i.e.\/}, (i) the magnitude distribution,
(ii) the local recurrence-time distribution, (iii) the global recurrence-time 
distribution, (iv) the time evolution of the spatial seismic distribution 
before the mainshock,
(v) the time evolution of the spatial seismic distribution 
after the mainshock, and (vi) the 
time-resolved magnitude distribution before and after the mainshock.
We show these results consecutively below.

\begin{figure}[ht]
\begin{center}
\includegraphics[scale=0.65]{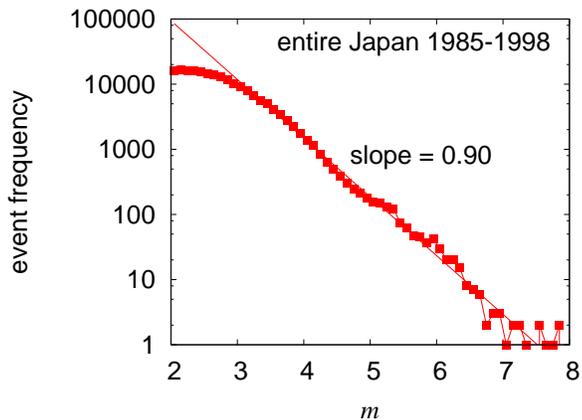}
\end{center}
\baselineskip 3.2mm
\caption{\small 
The magnitude distribution of earthquakes in Japan generated from 
the JUNEC catalog. The data for $m>3$ lie on a straight line with the
GR exponent $b\simeq 0.9$.
}
\label{HK:fig2}
\end{figure}

\subsection{The magnitude distribution}

In Fig.\ref{HK:fig2} we show the magnitude
distribution $R(m)$ of earthquakes of Japan
generated from the JUNEC catalog, 
where $R(m){\rm d}m$ represents the rate of events with their
magnitudes in the range [$m$, $m$ +{\rm d}$m$]. The 
data lie on a straight line fairly well for $m>3$, obeying the
GR law. The slope of the straight line gives the power-law exponent
about $b\simeq 0.9$.

\begin{figure}[ht]
\begin{center}
\includegraphics[scale=0.65]{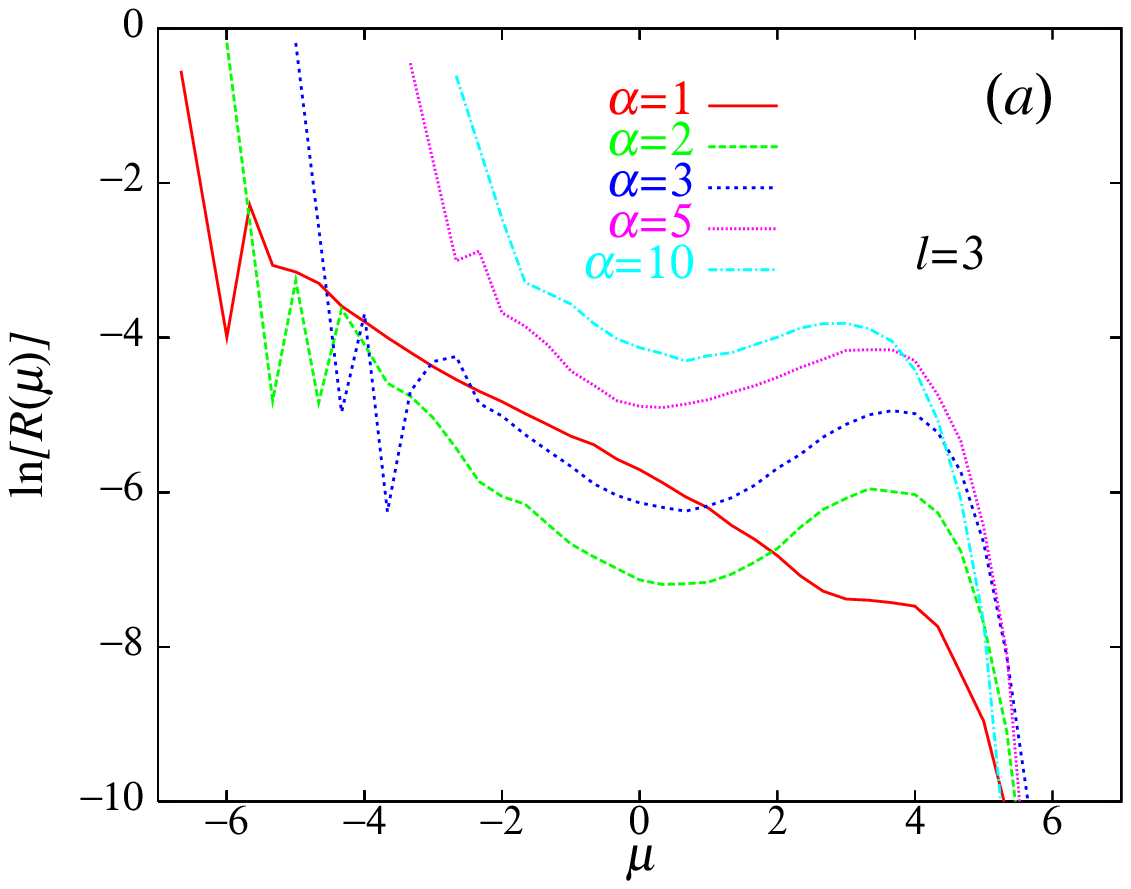}
\includegraphics[scale=0.65]{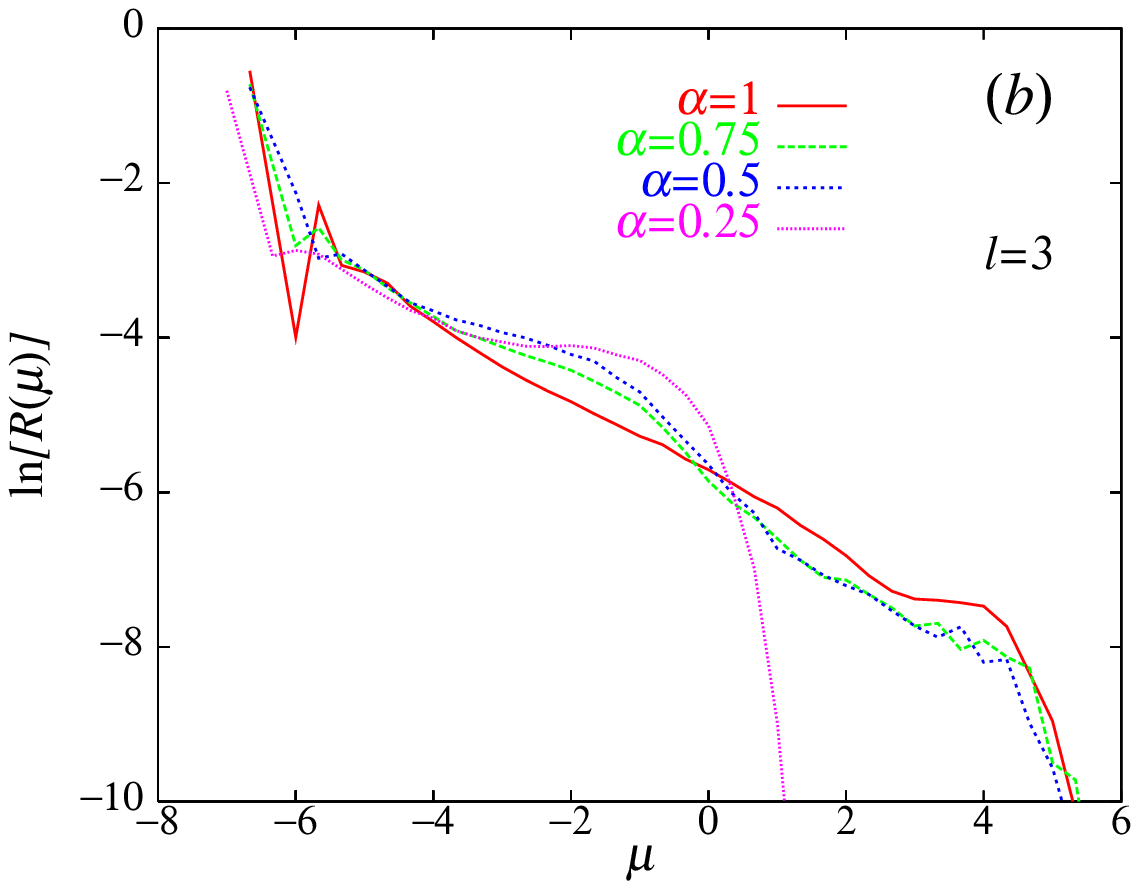}
\end{center}
\baselineskip 3.2mm
\caption{\small 
The magnitude distribution of seismic events
calculated for the 1D BK model, for the range of larger $\alpha\geq 1$ (a), 
and for the range of smaller $\alpha\leq 1$ (b).
}
\label{HK:fig3}
\end{figure}

In Fig.\ref{HK:fig3}(a), we show the magnitude distribution 
$R(\mu)$ of earthquakes
calculated from our numerical simulation of the 1D BK model. 
The parameter $\alpha$ is varied in the range
$0.25 \leq \alpha\leq 10$, while the elastic parameter $l$ 
is fixed to $l=3$.

In the BK model, the magnitude of an event, $\mu$, 
is defined as the logarithm of the moment $M_0$, {\it i.e.\/}, 
\begin{equation}
\mu=\ln M_0, \ \ \ M_0=\sum_i \Delta u_i,
\end{equation}
where $\Delta u_i$ is the 
total displacement of the $i$-th block during a given 
event and the sum is taken over all blocks involved in the event
\cite{HK:CLST}. It should be noticed that
the absolute numerics of the magnitude value of the
BK model $\mu$ has no direct quantitative correspondence to the
magnitude $m$ 
of real earthquakes.

\begin{figure}[ht]
\begin{center}
\includegraphics[scale=0.65]{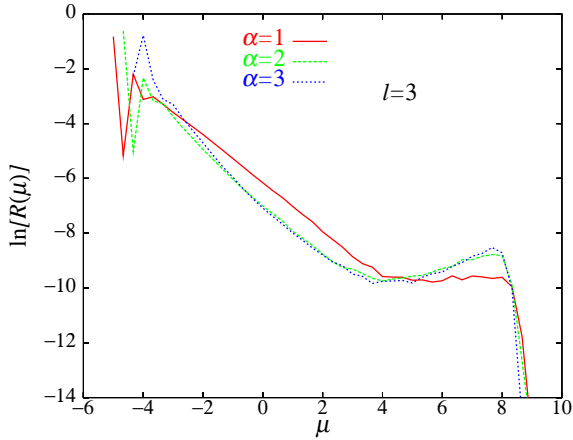}
\end{center}
\baselineskip 3.2mm
\caption{\small 
The magnitude distribution of seismic events
calculated for the 2D BK model.
}
\label{HK:fig4}
\end{figure}

As can be seen from Fig.\ref{HK:fig3}, the form of the calculated 
magnitude distribution depends on the
$\alpha$-value considerably.
The data  for  $\alpha =1$ lie on a straight line fairly well, apparently
satisfying the GR law. The value of the exponent $B$ describing 
the GR-like power-law behavior, 
$\propto 10^{-B}$, is estimated to be $B\simeq 0.50$ corresponding to
$b\simeq 0.75$, which is slightly smaller than the $b$-value observed for the
JUNEC catalog $b\simeq 0.9$. Remember the relation $b=\frac{3}{2}B$.

On the other hand,
the data for larger $\alpha$, {\it i.e.\/}, the ones for $\alpha \geq 2$
deviate
from the GR law at larger magnitudes, exhibiting a pronounced peak structure, 
while the power-law feature still remains for smaller magnitudes.
These features of the magnitude distribution are consistent with the
earlier observation of Carlson and Langer \cite{HK:CLST,HK:CL}. 
It means that,
while smaller events 
exhibit self-similar critical properties,  larger events tend to exhibit
off-critical or characteristic properties, much more so as
the velocity-weakening tendency of the friction  is increased.
The observed peak structure 
gives us a criterion to distinguish large and small events. Below,
we regard events with their magnitudes $\mu$ greater than $\mu_c=3$
as large events of the 1D BK model, $\mu_c=3$ being 
close to the peak position of the magnitude distribution 
of Fig.\ref{HK:fig3}(a). 
In an earthquake with  $\mu=3$,
the mean number of moving blocks are about 76 ($\alpha =1$) and 60 
($\alpha =2,3$).

As can be seen from Fig.\ref{HK:fig3}(b), the data at smaller $\alpha <1$
exhibit considerably different behaviors from those for $\alpha >1$.
Large events are suppressed here. For $\alpha =0.25$, in particular, 
all events consist almost exclusively of small events only. 
This result might be consistent with the earlier observation
which suggested that 
the smaller value of $\alpha <1$ tended to cause 
a creeping-like behavior without a large event \cite{HK:CL}.
In particular, Vasconcelos showed that a single block system 
exhibited a ``first-order
transition'' at $\alpha =0.5$ from a stick-slip to 
a creep \cite{HK:Vasconcelos},
whereas this
discontinuous transition becomes apparently continuous in many-block system
\cite{HK:Vieira,HK:Clancy}.
Since  we are mostly
interested in large seismic events here,
we concentrate  in the following on the parameter range
$\alpha \geq 1$. 
In contrast to the parameter $\alpha$,
the magnitude distribution turns out to be
less sensitive to the stiffness parameter
$l$. Further 
details of the $l$-dependence is given in Ref.\cite{HK:MoriKawamura2}.

In Fig.\ref{HK:fig4}, we show the magnitude 
distribution $R(\mu)$ of the 2D BK model with
varying the $\alpha$ value. At larger magnitudes, a deviation from
the GR law is observed for any value of $\alpha$, {\it i.e.\/},
the calculated magnitude distribution
exhibits a peak structure at larger magnitude irrespective of the
$\alpha$-value, suggesting that 
larger earthquakes tend to be characteristic.
From the observed peak structure, 
we regard events with their magnitudes $\mu$ greater than $\mu_c=5$
as large events of the 2D BK model.

While the BK model tends to reproduce the GR law  for earthquakes
of smaller magnitudes, its relevance to the GR law observed in 
real seimicity is not entirely clear. It has been suggested that
the GR $b$-value might be related to the fractal dimension of the fault
interface \cite{HK:Hanks,HK:Andrews,HK:Aki}.
More recently, Chakrabarti and collaborators proposed an earthquake model
in which the magnitude
distribution of earthquakes
is related to the contact area distribution between the 
two fractal surfaces of the plates \cite{HK:Chakarabarti}.
In these approaches, the intrinsic nonuniformity at earthquake faults 
plays an essential role in realizing the GR law and the 
critical nature of earthquakes. Whether this is really true, as well as 
its possible relation to the BK model where the nonuniformity 
is apparently absent, or at least not explicit, needs to be examined further.

\subsection{The local recurrence-time distribution}

A question of general interest 
may be how large earthquakes repeat in time, 
do they occur near periodically or irregularly? 
One may ask this question either locally, {\it i.e.\/}, for a given finite 
area on the fault,
or globally,  {\it i.e.\/}, for an entire fault system.
The picture of characteristic
earthquake presumes the existence of a characteristic recurrence time. 
In this case, the distribution of the recurrence time of large earthquakes, 
$T$, 
is expected to exhibit a peak structure at such a characteristic time scale.
If the SOC concept applies to large earthquakes, by contrast,
such a peak structure would not show up. 

\begin{figure}[]
\begin{center}
\includegraphics[scale=0.65]{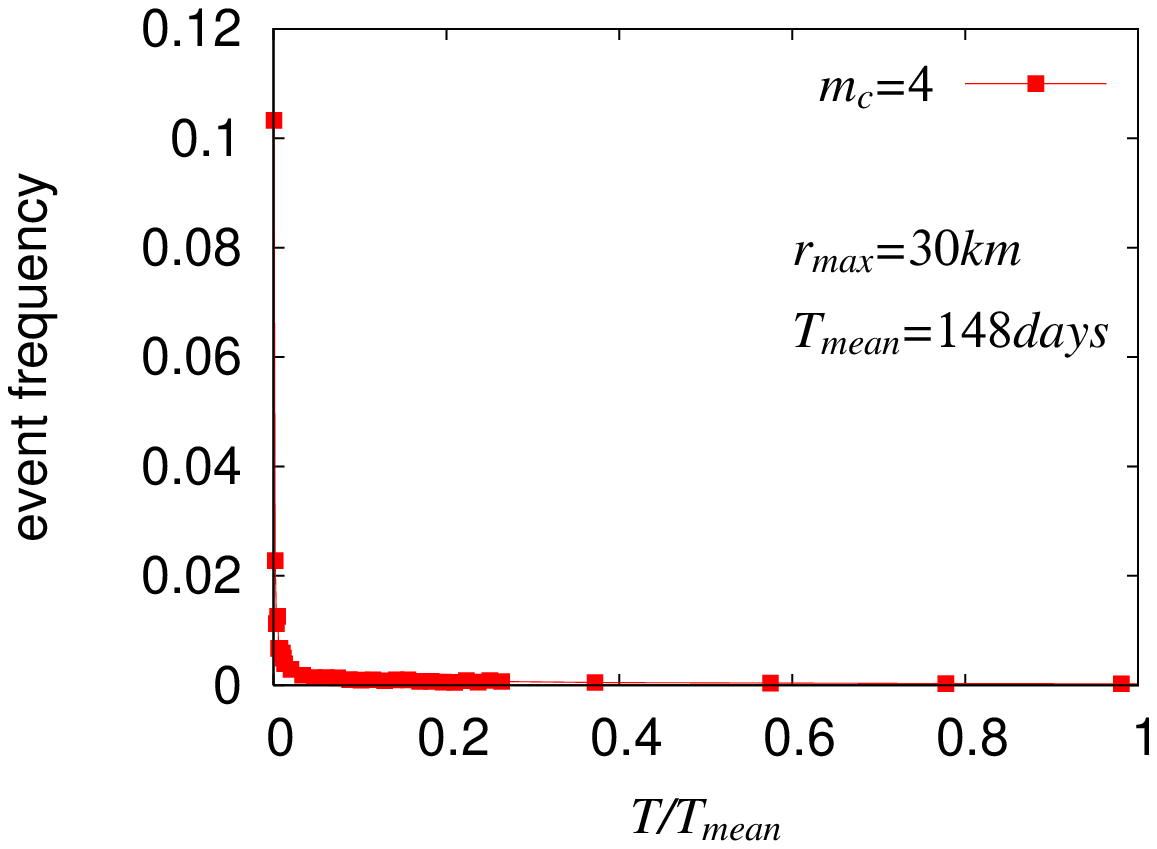}
\includegraphics[scale=0.65]{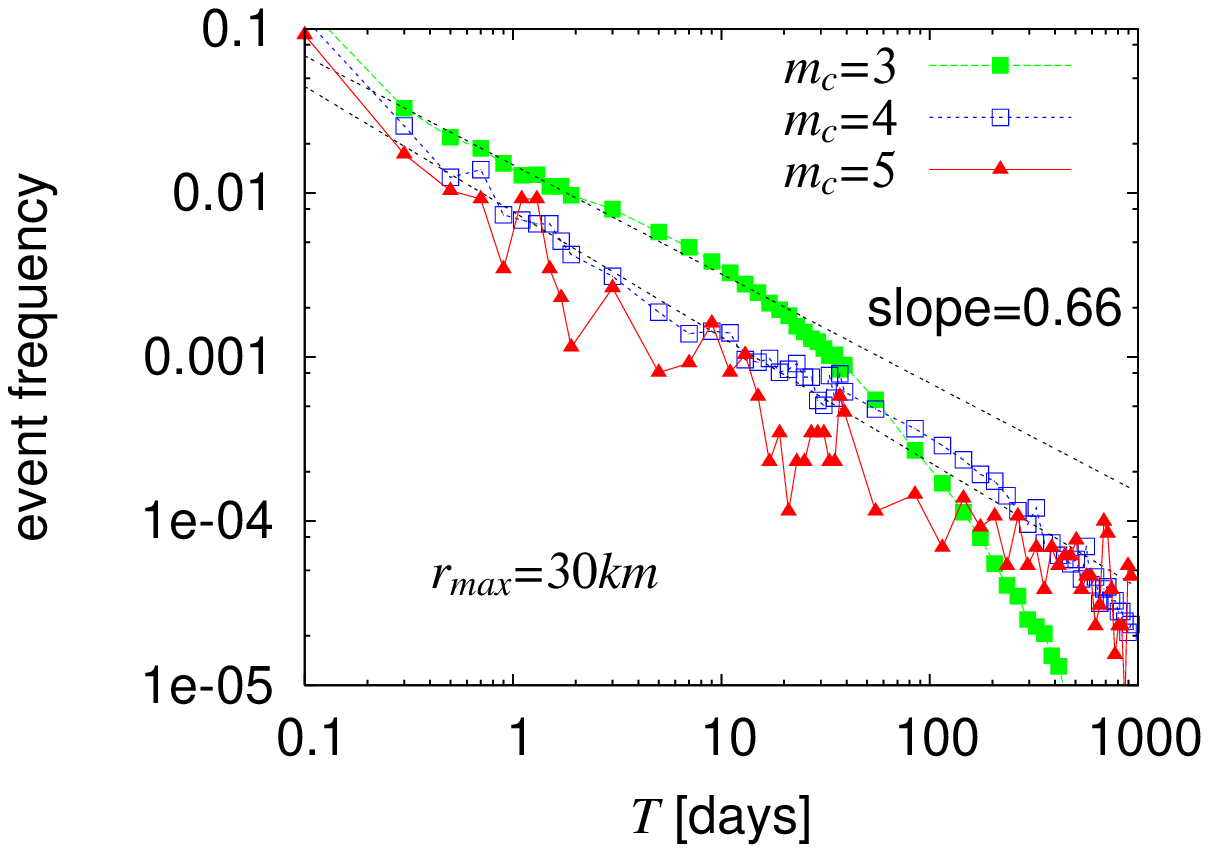}
\includegraphics[scale=0.65]{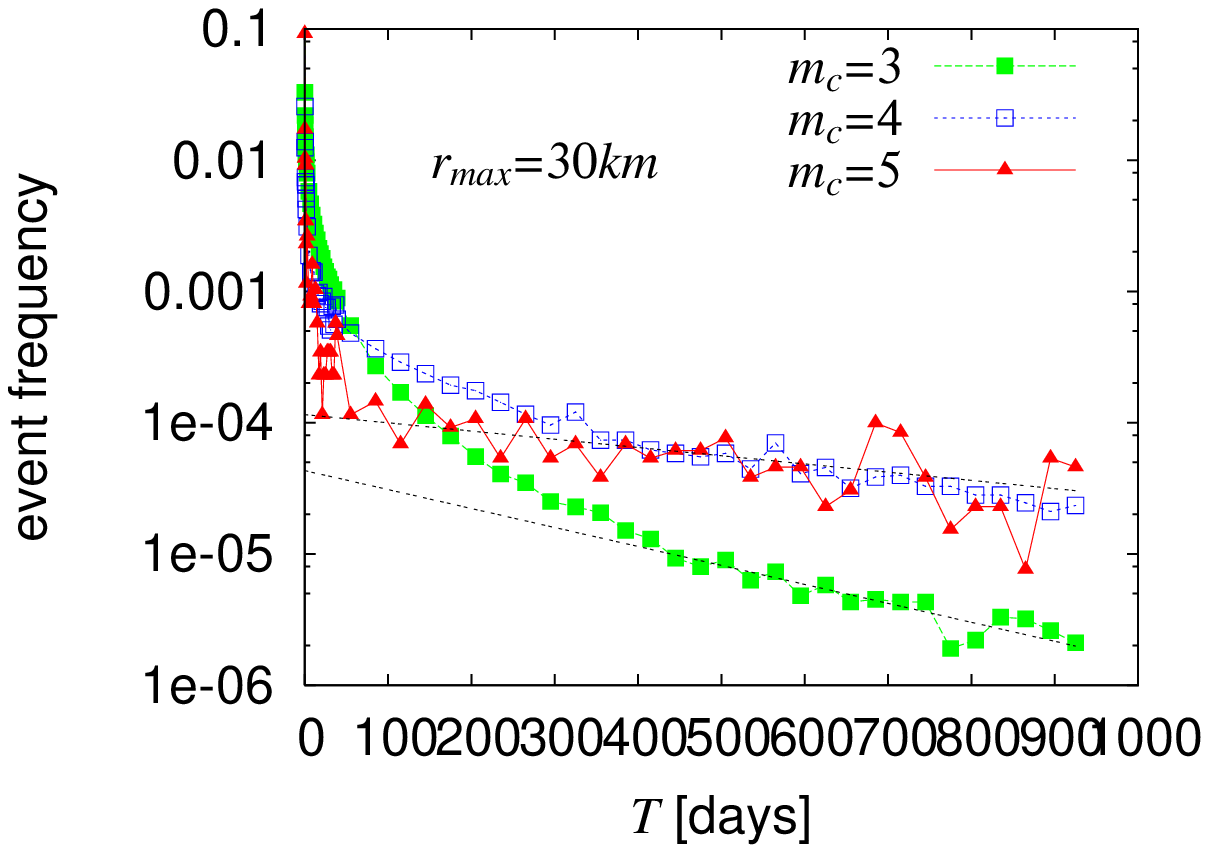}
\end{center}
\baselineskip 3.2mm
\caption{\small 
The local recurrence-time distribution of large earthquakes in Japan
generated from the JUNEC catalog. The distribution is given on a bare scale 
for large 
earthquakes with $m_c=4$ [upper figure], on a logarithmic
scale with $m_c=3,4$ and 5 [middle figure], and on a semi-logarithmic scale
with  $m_c=3,4$ and 5 [lower figure].
}
\label{HK:fig5}
\end{figure}

In the upper panel of Fig.\ref{HK:fig5}, we show the distribution of the local
recurrence time $T$ of large earthquakes of Japan
with $m \geq m_c=4$, 
calculated from the JUNEC catalog. 
In defining the recurrence time
locally, 
the subsequent large event is counted when a large
event with $m\geq m_c=4$ occurs 
with its epicenter lying within a circle of radius 30 km
centered at the epicenter of the previous large event.
The mean recurrence time $\bar T$ is then estimated to be 148 days.
In the middle and lower panels of Figs.\ref{HK:fig5}, 
the same data are re-plotted on a double-logarithmic 
scale [middle panel], and on a semi-logarithmic scale [lower panel],
with the magnitude threshold $m_c=3,4$ and 5.
One sees from these figures that the local recurrence-time
distribution exhibits  power-law-like critical features at
shorter times, which seems to cross over to a faster exponential-like decay at
longer times. The time range in which the data obey a power-law
tends to get longer for larger earthquakes. The power-law
exponent estimated for earthquakes of $m_c=5$ in the
time range $T<1000$ days is about $\frac{2}{3}$. This value is
a bit smaller than the standard Omori exponent $\simeq 1$.

\begin{figure}[htb]
\begin{center}
\includegraphics[scale=0.65]{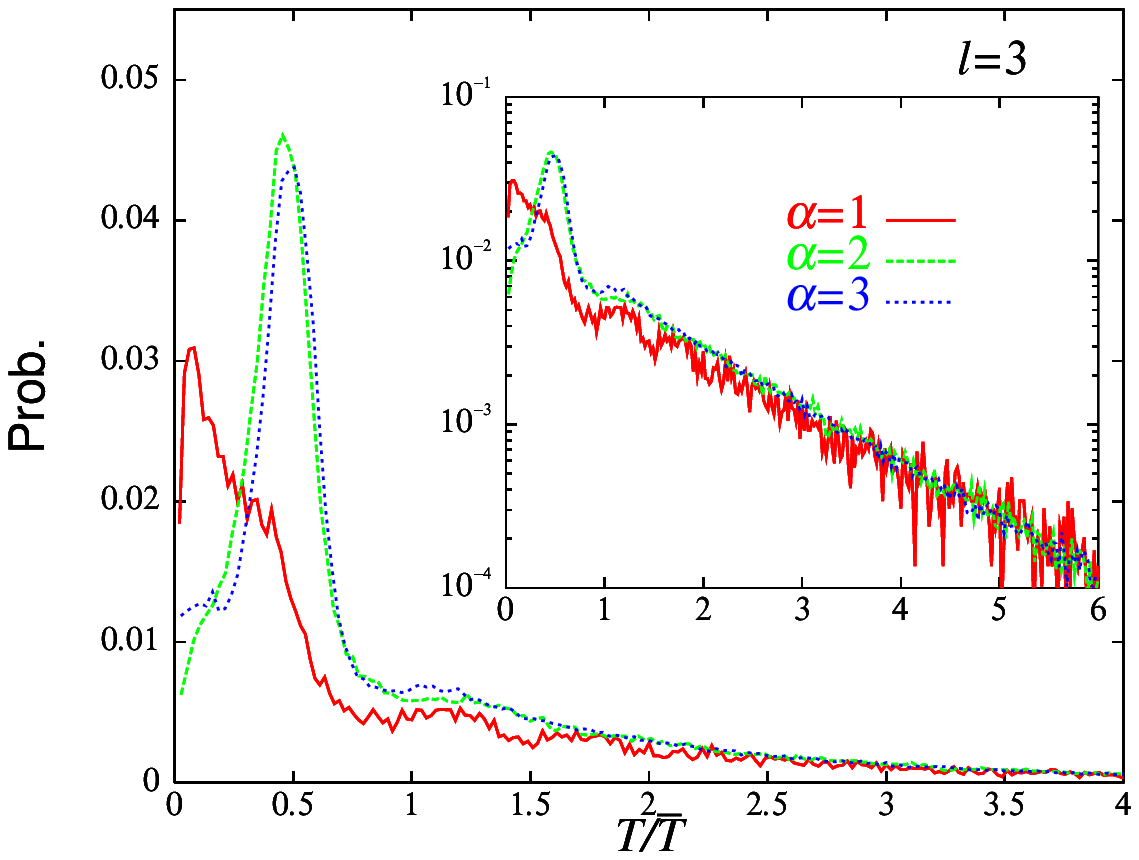}
\includegraphics[scale=0.65]{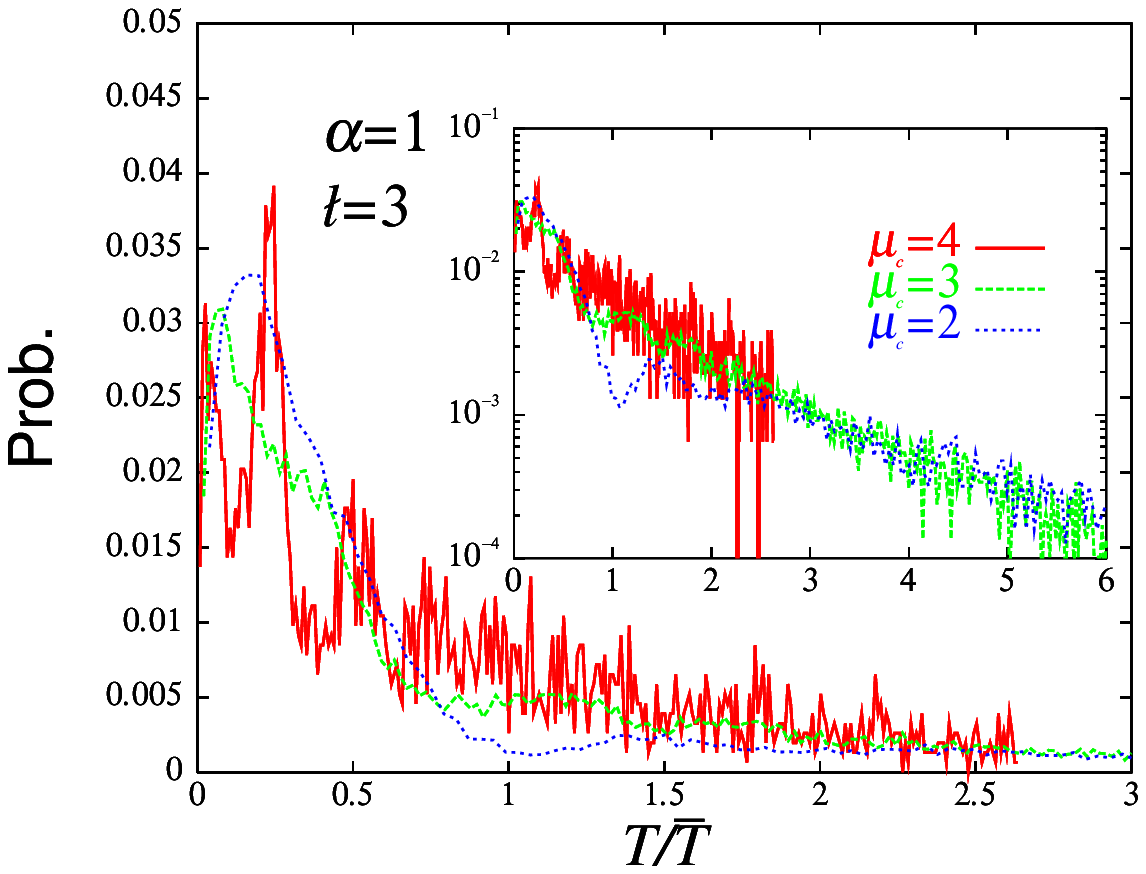}
\end{center}
\baselineskip 3.2mm
\caption{\small 
The local recurrence-time distribution of large events of the 1D BK model. In the
upper figure, the distribution is given for $\alpha=1,2$ and 3 with fixing
$\mu_c=3$, whereas, in the
lower figure, the distribution is given with varying the magnitude threshold
as $\mu_c=2,3$ and 4 with fixing $\alpha=1$.
}
\label{HK:fig6}
\end{figure}

In the upper panel of Fig.\ref{HK:fig6}, we show the distribution of the
local recurrence time $T$ of large earthquakes with $\mu \geq \mu _c=3$, 
calculated for the 1D BK model.
In the insets, the same data including the tail part 
are re-plotted on a 
semi-logarithmic scale.
In defining the recurrence time locally in the BK model, 
the subsequent large event is counted when a large
event occurs with its epicenter in the region within
30 blocks from the epicenter of the previous large event.
The mean recurrence time $\bar T$ is then estimated to be
$\bar T\nu =1.47$, 1.12, and 1.13 for $\alpha=1$, 2 and 3,  respectively.

As can be seen from Fig.\ref{HK:fig6},
the tail of the distribution  is exponential at longer $T> \bar T$
for all values of $\alpha$, 
while the form of the distribution at shorter $T< \bar T$ 
is non-exponential 
and differs between
for $\alpha =1$ and for $\alpha =2$ and 3. 
For $\alpha=2$ and 3, the distribution
has an eminent  peak at around
$\bar T\nu \simeq 0.5$, not far from the mean  
recurrence time.
This means the existence of a characteristic recurrence time,
suggesting the near-periodic recurrence of 
large events. This characteristic behavior is in sharp contrast to the
critical behavior without any peak structure observed in the JUNEC data.
It should also be mentioned, however, that there were reports in the literature
of a near-periodic recurrence of large
events at several real faults \cite{HK:ScholzBook,HK:NB}.

For $\alpha =1$, by contrast,
the peak located close to the mean $\bar T$ is hardly discernible. 
Instead, the distribution has a pronounced peak at a 
shorter time $\bar T\nu \simeq 0.10$, just after the previous large event.
In other words, large events for $\alpha=1$ tend to occur as 
 ``twins''. This has also been confirmed by our analysis of the time record
of large events. In fact, as shown in Ref.\cite{HK:MoriKawamura2}, 
a large event for the case 
of $\alpha=1$ often occurs as a ``unilateral earthquake'' 
where the rupture  propagates
only in one direction, hardly propagating 
in the other direction.
When a large earthquake occurs in the form of  
such a unilateral earthquake, further
loading due to the plate motion tends to trigger the subsequent
large event in the opposite direction, causing a twin-like event. 
This naturally explains the small-$T$ peak observed in Figs.\ref{HK:fig6} 
for $\alpha=1$.
\begin{figure}[ht]
\begin{center}
\includegraphics[scale=0.65]{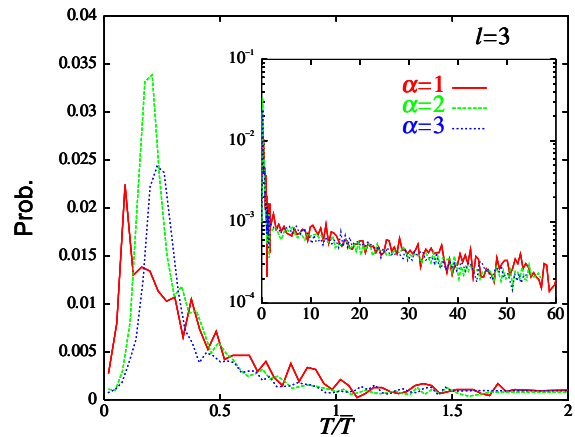}
\end{center}
\baselineskip 3.2mm
\caption{\small 
The local recurrence-time distribution of large events 
of $\mu > \mu_c=5$ of the 2D BK model. 
}
\label{HK:fig7}
\end{figure}

In the lower panel of Fig.\ref{HK:fig6}, the local
recurrence-time distribution of large events is shown for the cases of 
$\alpha =1$, with varying the 
magnitude threshold as $\mu _c=2$, 3 and 4. As can be seen form this figure,
the form of the
distribution for
$\alpha=1$ largely changes with the threshold value $\mu_c$.
Interestingly, in the case of  $\mu_c=4$, 
the distribution has {\it two\/} 
distinct peaks, one corresponding to 
the twin-like event and the other to the near-periodic event.
Thus, even in the case of $\alpha=1$ where the critical features
are apparently dominant for smaller thresholds $\mu_c=2$ and 3, features of 
a characteristic earthquake becomes 
increasingly eminent when 
one looks at very large events.

In Fig.\ref{HK:fig7}, the local recurrence-time distribution is shown for the 2D BK model. 
In defining the recurrence time locally in the 2D BK model, 
a subsequent large event is counted when a large
event occurs with its epicenter lying within a circle of its radius
5 blocks centered at the epicenter of the previous large event.
The mean recurrence time $\bar T$ is then estimated to be
$\bar T\nu =1.47$, 1.12, and 1.13 for $\alpha=1$, 2 and 3,  respectively.
The behavior of the 2D model is similar to the 1D model, with enhanced 
characteristic features. The peak structure of the distribution is 
very prominent even for $\alpha=1$.

\subsection{The global recurrence-time distribution}

In this subsection, we examine the global recurrence-time distribution.
By the word ``global'', we consider the situation where the region in which
one identifies the next event is sufficiently wide, much wider than the 
typical size of the rupture zone of large events.

\begin{figure}[]
\begin{center}
\includegraphics[scale=0.65]{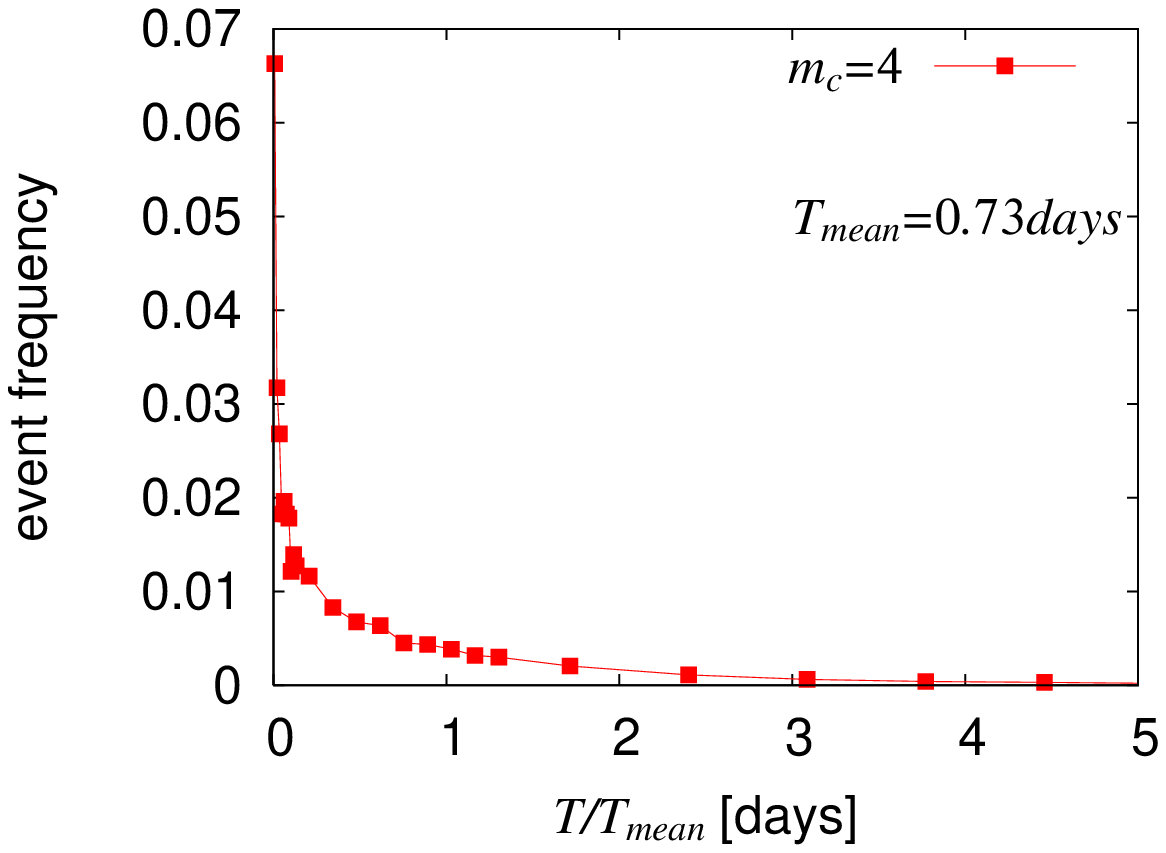}
\includegraphics[scale=0.65]{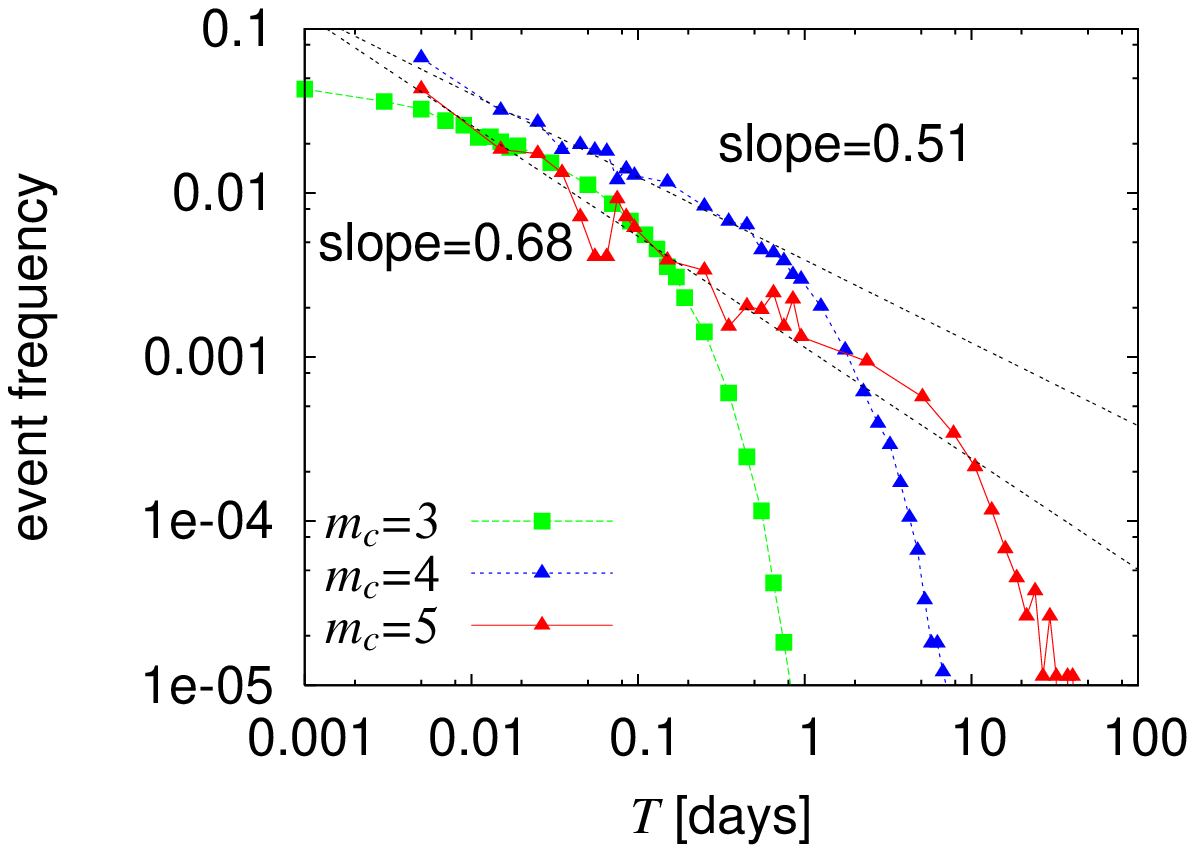}
\includegraphics[scale=0.65]{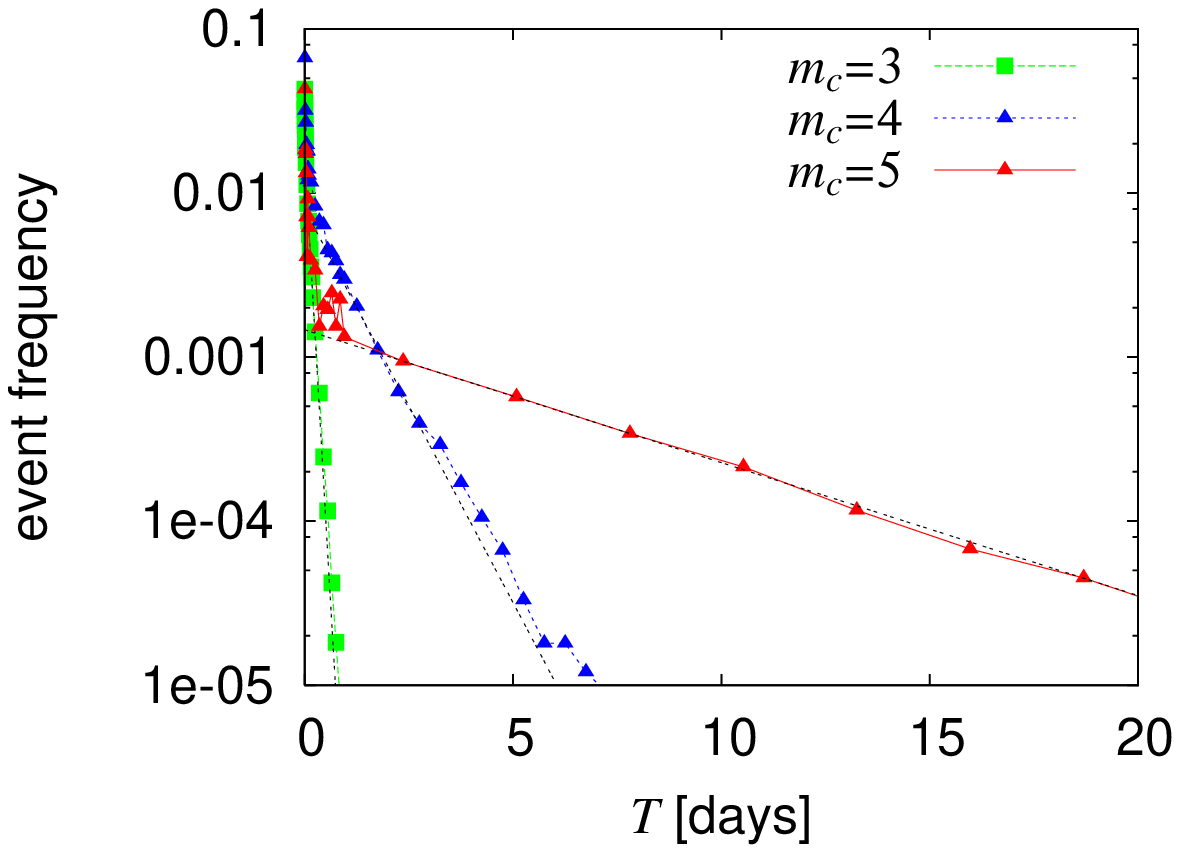}
\end{center}
\baselineskip 3.2mm
\caption{\small 
The global recurrence-time distribution of large earthquakes in entire Japan
generated from the JUNEC catalog. The distribution is given on a bare scale 
for large 
earthquakes with $m_c=4$ [upper figure], on a double-logarithmic
scale with varying $m_c=3,4$ and 5 [middle figure], 
and on a semi-logarithmic scale
with varying $m_c=3,4$ and 5 [lower figure].
}
\label{HK:fig8}
\end{figure}

From the JUNEC catalog, we construct in the upper panel of 
Fig.\ref{HK:fig8} such
{\it global\/} recurrence-time distribution for entire Japan
for large earthquakes with $m_c=4$.
The mean recurrence time $\bar T$ is then estimated to be 0.73 days.
In the middle and lower panels of Figs.\ref{HK:fig8}, 
the same data are re-plotted on a 
double-logarithmic scale [middle], and on a semi-logarithmic scale 
[lower],
with the magnitude threshold $m_c=3,4$ and 5.
One sees from these figures that the global recurrence-time
distribution of the JUNEC data is very much similar to the corresponding
local recurrence-time
distribution shown in Figs.\ref{HK:fig5}: 
It exhibits  power-law-like critical features at
shorter times, which crosses over to a faster exponential-like decay at
longer times. The time range in which the data obey a power-law
gets longer for larger earthquakes. The
exponent describing the power-law regime
is roughly about $\frac{2}{3}$, which is not far from the
corresponding exponent of the local recurrence-time distribution. 

\begin{figure}[ht]
\begin{center}
\includegraphics[scale=0.65]{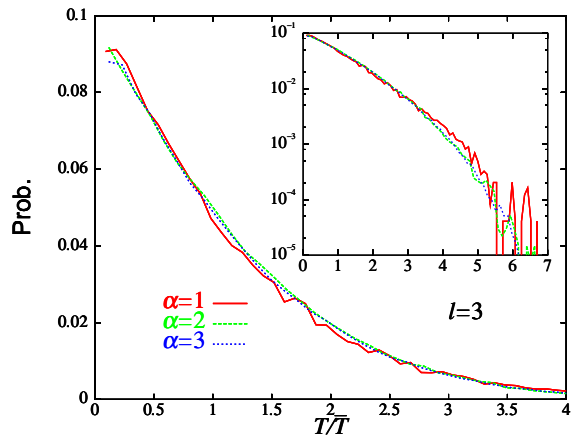}
\end{center}
\baselineskip 3.2mm
\caption{\small 
The global recurrence-time distribution of large events of the 1D BK model. 
The distribution is given for $\alpha=1,2$ and 3 with fixing the magnitude
threshold $\mu_c=3$.
}
\label{HK:fig9}
\end{figure}

Fig.\ref{HK:fig9} exhibits 
the {\it global\/} recurrence-time distribution of large events
with $\mu_c=3$ calculated for the 1D BK model.
As can clearly be seen from the figure, in the BK model, the form of the
distribution takes a different form from the local one:
The peak structure seen in the local distribution no longer 
exists here. Furthermore, 
the form of the distribution tail at larger $T$
is not a simple exponential, faster than exponential:
See a curvature of the data in the inset of Fig.\ref{HK:fig9}. 
These features of the global recurrence-time distribution turn out to be rather
robust against the change of the parameter values such as $\alpha $ and $l$, 
as long as the system size is taken to be sufficiently large.

The observation that the local and the global recurrence-time
distributions exhibit mutually different behaviors means that the form of the
distribution  depends on the length scale
of measurements. Such scale-dependent features of the
recurrence-time distribution of the BK model are 
in apparent contrast with the
scale-invariant power-law features of the
recurrence-time distributions observed in the JUNEC
data shown in Figs.\ref{HK:fig5} and \ref{HK:fig8},
and the ones
reported for some of real faults 
\cite{HK:Bak02,HK:Corral}. We note that essentially the same behavior 
as the one of the 1D BK model is also
observed in the 2D BK model (the data not shown here).

\subsection{Spatiotemporal seismic correlations before the mainshock}

In this subsection, we study the spatiotemporal correlations of 
earthquake events before the mainshock.

We begin with the JUNEC catalog.
First, we define somewhat arbitrarily the mainshock 
as an event whose magnitude $m$
is greater than $m_c=5$, and
pay attention to the frequency of earthquake
events preceding the mainshock, particularly its time  and distance
dependence. Both the time $t$ and the distance $r$ are measured 
relative to the
time and the position of the subsequent mainshock. 
The distance between the two events is measured here as the distance
between their epicenters. The event frequency is
normalized by the factor $r$ associated with the area element of the
polar coordinate.

\begin{figure}[ht]
\begin{center}
\includegraphics[scale=0.65]{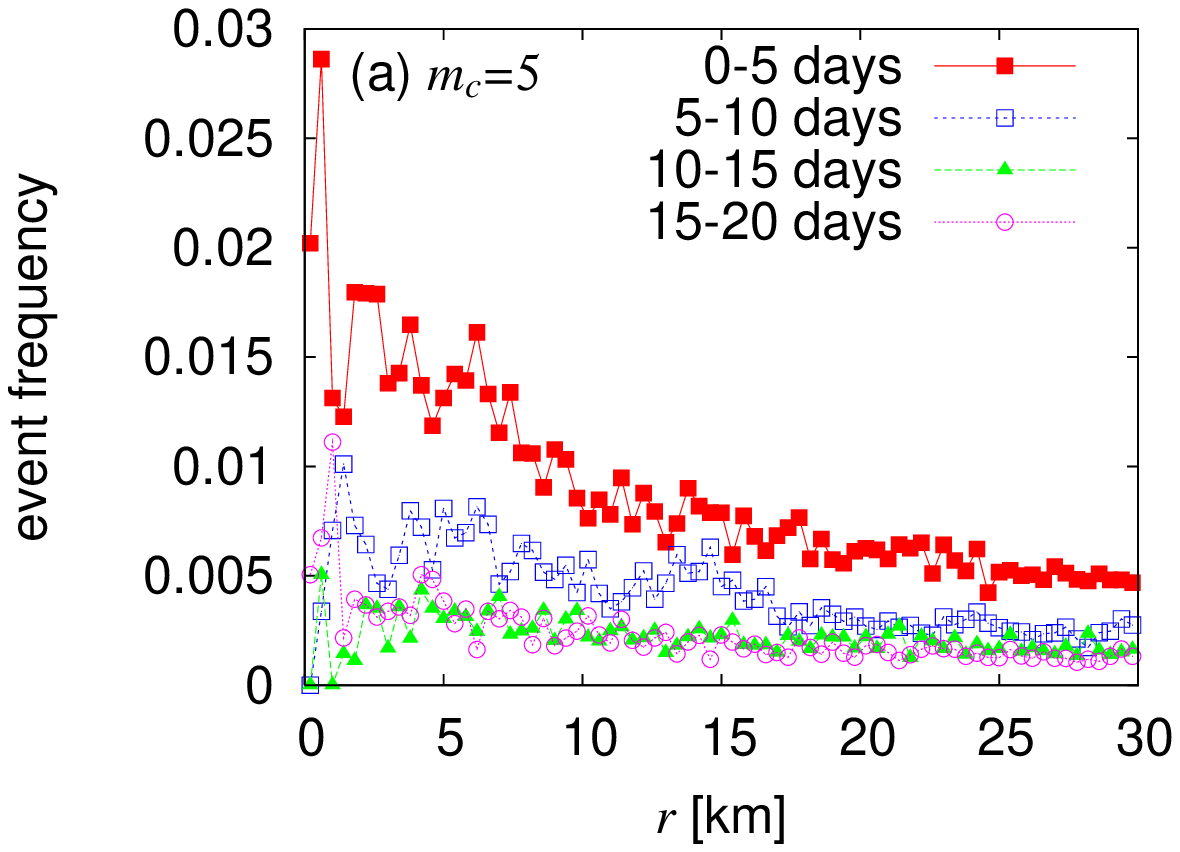}
\includegraphics[scale=0.65]{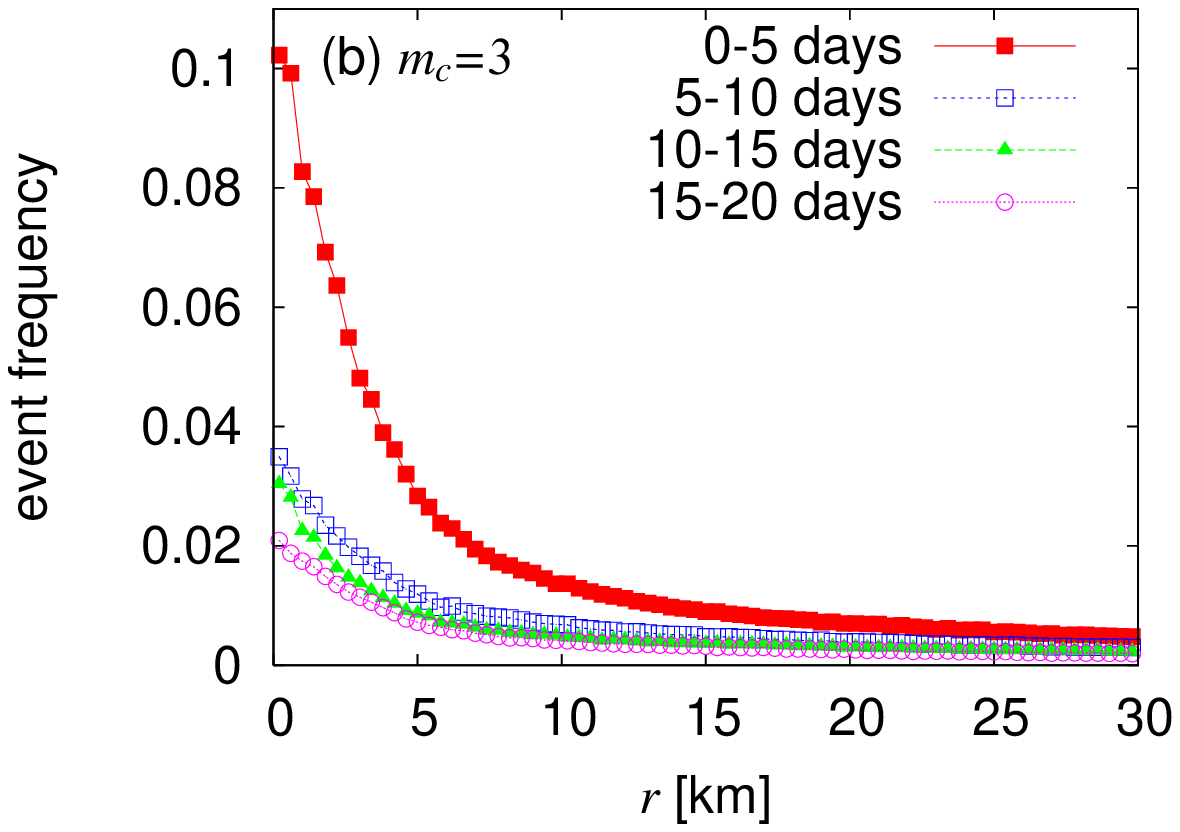}
\end{center}
\baselineskip 3.2mm
\caption{\small 
The time evolution of the event frequency of arbitrary magnitude preceding
the mainshock, 
plotted as a function of 
the distance from the epicenter of the upcoming mainshock. The magnitude 
threshold for the mainshock is $m_c=5$ (a) and $m_c=3$ (b).
The data are generated from the JUNEC catalog, while 
the contribution from the two special regions, {\it i.e.\/}, the Izu
region and the Ebino region, has been omitted from the data: See the text
for further details.
}
\label{HK:fig10}
\end{figure}

Fig.\ref{HK:fig10} exhibits the event frequency plotted as a function of $r$,
the distance from the upcoming mainshock, for four time periods preceding
the mainshock
each containing five days. The data show
the time evolution of the spatial pattern of event number irrespective
of its magnitude, 
which occur within
30 km from the epicenter of the upcoming mainshock and during the last
20 days toward the mainshock. The data have been 
averaged over the mainshocks contained in the data set 
taken over entire Japan, but {\it omitting
the contributions of the two special narrow regions, i.e., 
Izu and Ebino\/}. 
In these two regions, active earthquake ``swarms'' occurred during the period,
which lead to significantly different behavior in 
the spatiotemporal correlations (we return to this point later in this
subsection).
In the data set, total of 990 mainshocks
are included, each mainshock accompanying about 8.7 preceding events on average
in the time/space range shown in the figure.
As can clearly be seen from Fig.\ref{HK:fig10}(a), there is a tendency that the
seismic activity accelerates as the mainshock approaches, and this
tendency is 
more pronounced 
in a closer vicinity of the epicenter of the 
upcoming mainshock. The data demonstrate that the  mainshock accompanies 
{\it foreshocks\/}, at least in the statistical sense.

It should be emphasized here that the evidence of  foreshock activity
becomes clear only after averaging over a large number of mainshocks. 
Note that, in the time/space region of interest, 
each single mainshock accompanies  on average less than ten foreshocks
only, a too small number to say anything definite.

One may wonder if certain 
qualitative features of the spatiotemporal correlations
shown in 
Fig.\ref{HK:fig10}(a) might change depending on the region, the depth and the 
magnitude-threshold of the mainshock. 
Examination of the JUNEC data, however, 
reveals that the qualitative features of Fig.\ref{HK:fig10}(a) 
are rather robust against these
parameters. The average number of foreshocks depends somewhat on each region
in Japan: For example, the mean numbers of foreshocks 
in the space/time region of Fig.\ref{HK:fig10}(a) are 
9.4 for the northern part of Japan
(40$^\circ$N- latitude), 14.7 for the middle part (36$^\circ$N-40$^\circ$N)
and 3.6 for the southern part (-36$^\circ$N). Nevertheless, the 
qualitative features of the
spatiotemporal correlations turn out 
to be more or less common. 

In Fig.\ref{HK:fig10}(b), we show the similar
spatiotemporal seismic 
correlations as in Fig.\ref{HK:fig10}(a), 
but now 
reducing the magnitude threshold of the mainshock from $m_c=5$ to $m_c=3$.
By this definition, the total number of mainshocks is increased to 53,835 
so that the
better statistics is expected. Indeed, the data shown in Fig.\ref{HK:fig10}(b) 
are far
less erratic than those in Fig.\ref{HK:fig10}(a), 
reflecting the improvement of the statistics.
Nevertheless, the qualitative features remain almost the same as in 
Fig.\ref{HK:fig10}(a).

\begin{figure}[]
\begin{center}
\includegraphics[scale=0.65]{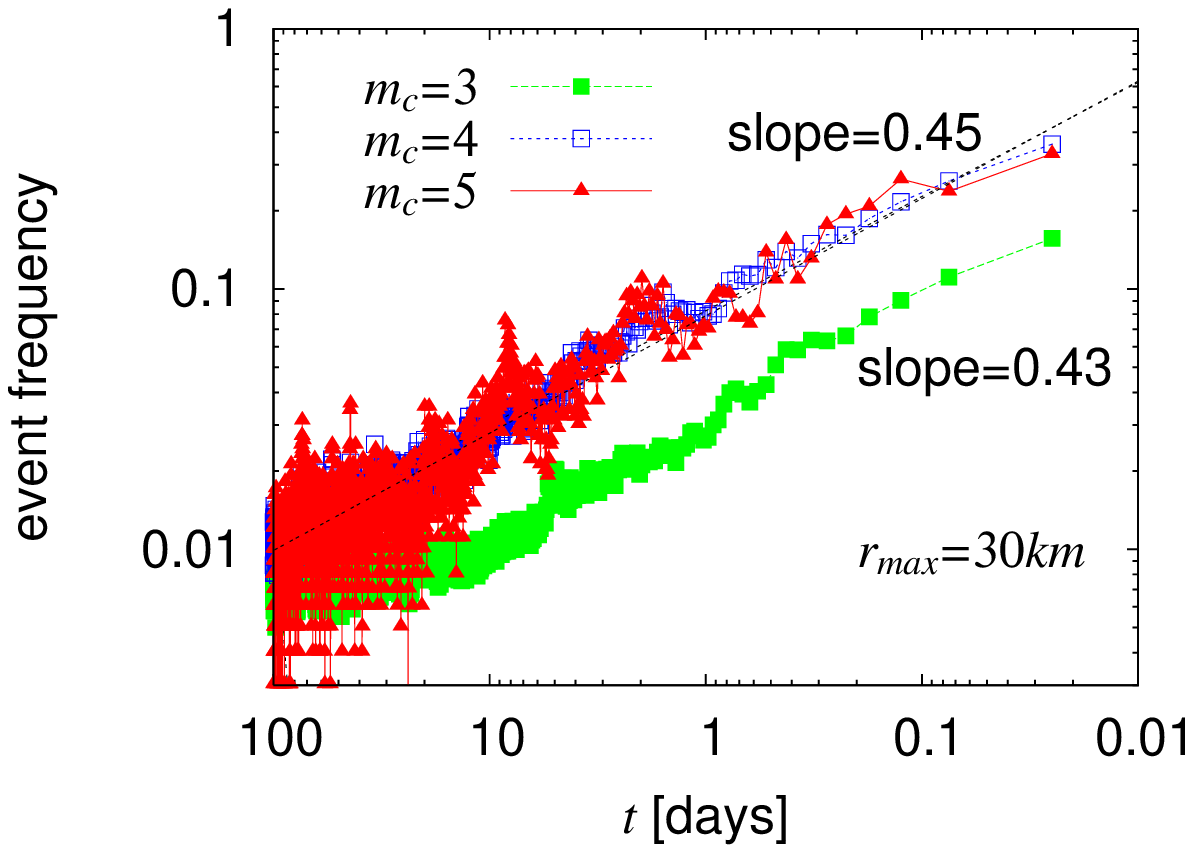}
\includegraphics[scale=0.65]{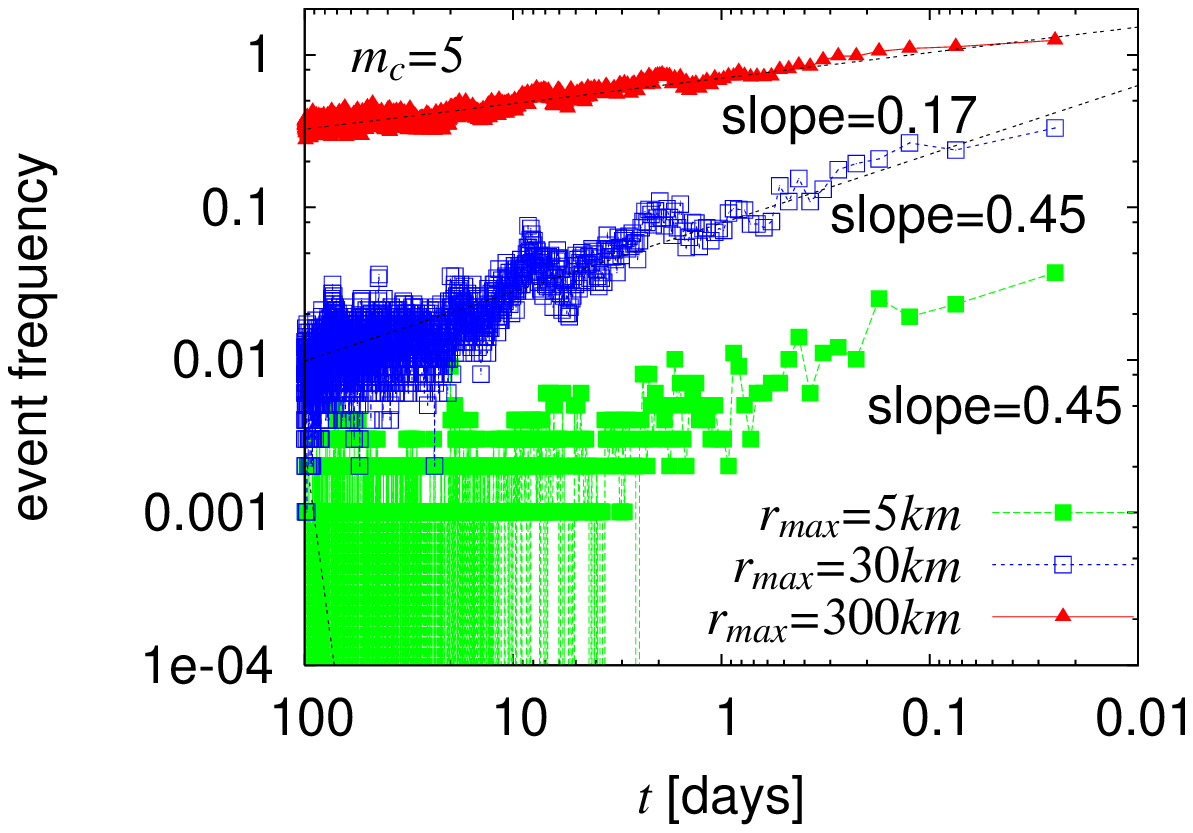}
\end{center}
\baselineskip 3.2mm
\caption{\small 
The frequency of seismic events of arbitrary magnitude preceding 
the mainshock, 
plotted versus the time until the mainshock on a double-logarithmic scale.
The data are generated from the JUNEC catalog, where
the contribution from the two special regions, {\it i.e.\/}, the Izu
region and the Ebino region, has been omitted: See the text
for further details.
In the upper figure, the magnitude threshold of the mainshock
is varied as $m_c=3,4$ and 5 with fixing the distance range of observation
$r_{max}=30$ km,
whereas, in the lower figure, the distance range of observation
is varied as $r_{max}=5,\ 30$ and 300 km
with fixing the magnitude threshold $m_c=5$.
}
\label{HK:fig11}
\end{figure}

In order to clarify the relation with the inverse Omori law which describes
the time evolution of the frequency of foreshocks, we show
in Fig.\ref{HK:fig11} 
the time dependence of the frequency of foreshocks before the mainshock
on a double-logarithmic scale: In the upper panel, 
the data with the distance range
$r_{max}=30$ km are shown with 
varying the magnitude threshold as $m_c=3,4$ and 5, while in the lower
panel, the data
with $m_c=5$ are shown  with varying the distance range 
as $r_{max}=5, 30$ and 300 km. 
Except for the case of very large distance range
$r_{max}=300$ km, the inverse Omori exponent comes around  
0.5.

Now, we wish to turn to the spatiotemporal correlations in the 
two special regions, the contribution of which have intentionally been omitted
in our analysis of Figs.\ref{HK:fig10} and \ref{HK:fig11}. 
In fact, these special regions are associated
with ``swarms''. If there occurs an earthquake swarm which contains in itself 
a few large events which can be regarded as mainshocks, 
they make a significant contribution to the above spatiotemporal correlations
because a huge number of small events are
contained in swarms. The two swarms we discarded in our analysis of 
Figs.\ref{HK:fig10} and \ref{HK:fig11}
are, (i) the Izu earthquake swarm, and (ii) the Ebino earthquake swarm 
(Ebino is located close to the Kirishima volcanoes in southern Kyusyu).
Indeed, if the contribution of these two swarm regions were not separated 
in our analysis of Figs.\ref{HK:fig10} and \ref{HK:fig11},
the resulting distribution function would look 
considerably different.

\begin{figure}[ht]
\begin{center}
\includegraphics[scale=0.65]{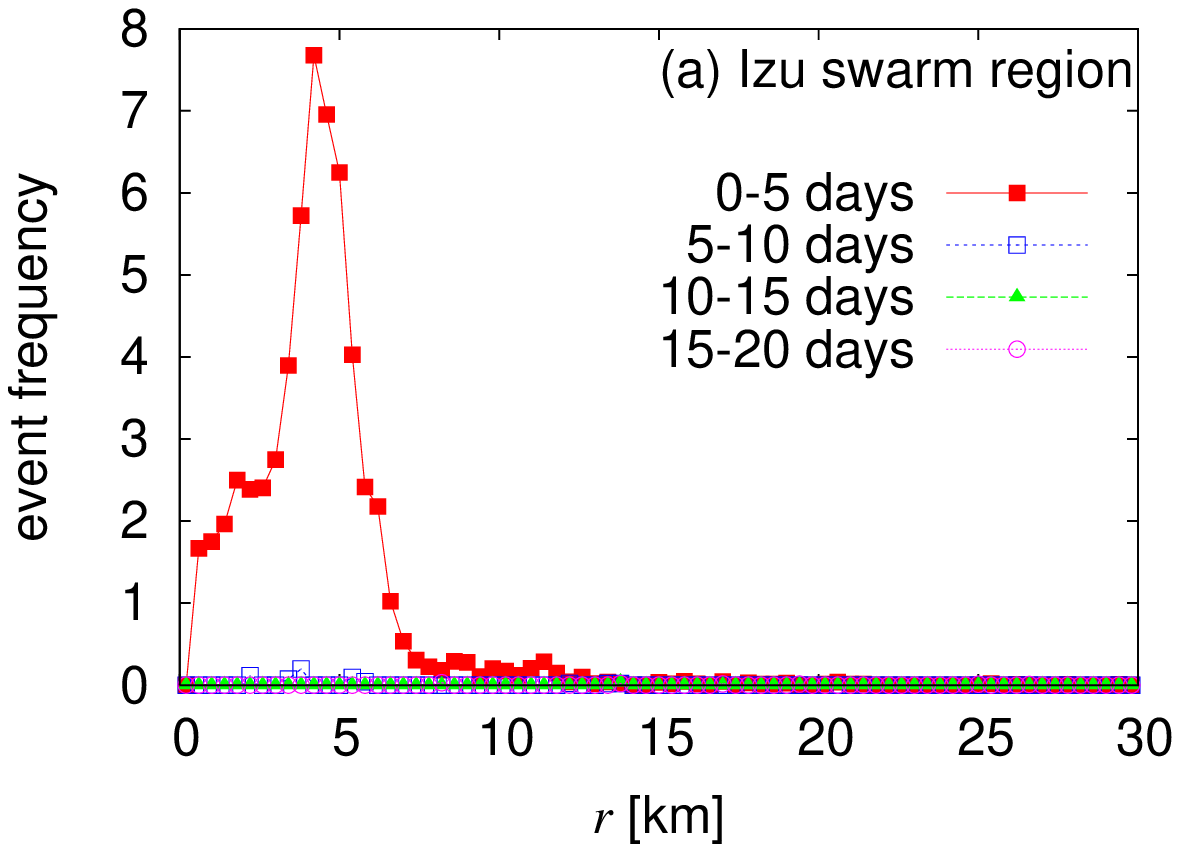}
\includegraphics[scale=0.65]{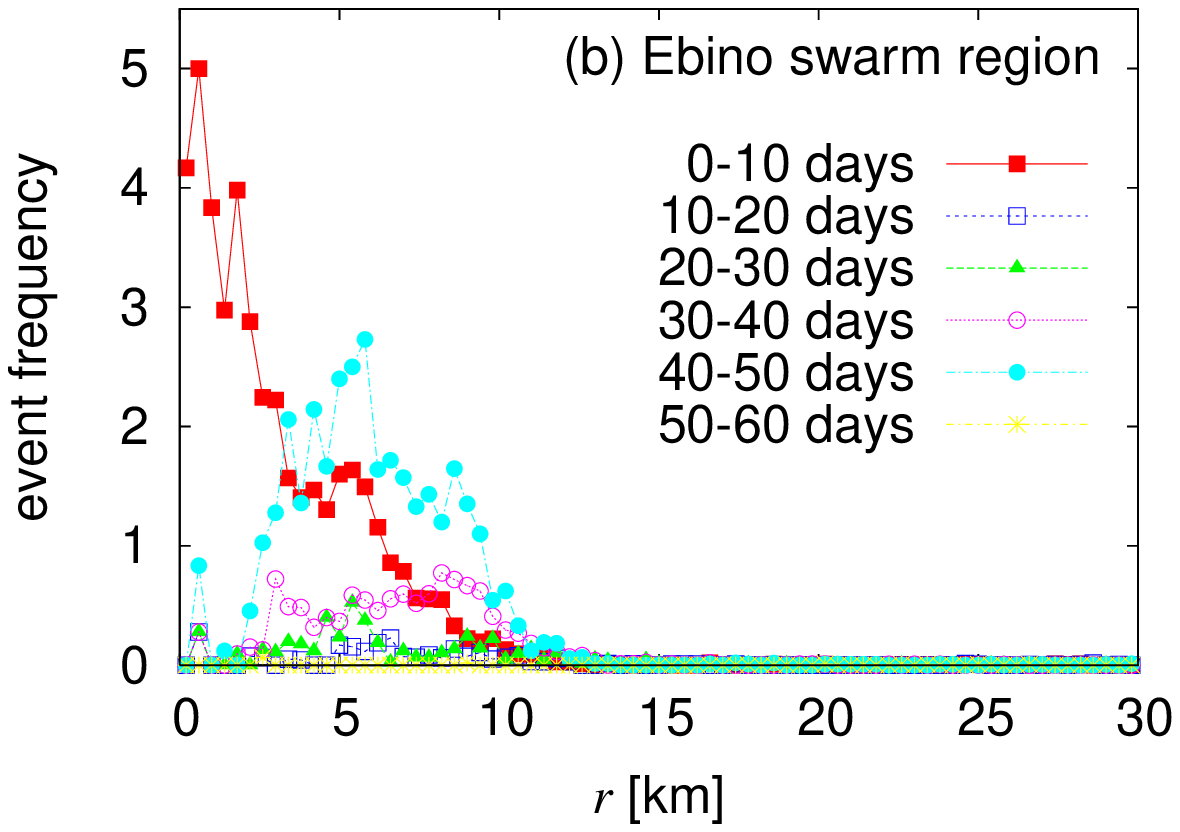}
\end{center}
\baselineskip 3.2mm
\caption{\small 
The time evolution of the event frequency of arbitrary magnitude  preceding
the mainshock,
plotted versus 
the distance from the epicenter of the upcoming mainshock. 
The data are generated from the JUNEC catalog,
for the Izu earthquake swarm region  
[34.8$^\circ$N-35.1$^\circ$N latitude; 139.0$^\circ$E-139.5$^\circ$E 
longitude] (a), and for the Ebino earthquake swarm region  [31.8$^\circ$N-32.0$^\circ$N; 130.2$^\circ$E-130.6$^\circ$E] 
(b), both during the period 1985-1998.  The magnitude threshold of the 
mainshock is $m >m_c=5$
}
\label{HK:fig12}
\end{figure}

In Figs.\ref{HK:fig12}, we show the the spatiotemporal seismic correlations
associated with mainshocks with $m_c=5$
in the Izu swarm region [34.8$^\circ$N-35.1$^\circ$N latitude; 
139.0$^\circ$E-139.5$^\circ$E longitude] (a),
and in the Ebino swarm region [31.8$^\circ$N-32.0$^\circ$N; 130.2$^\circ$E-130.6$^\circ$E] (b), respectively.
The total numbers of mainshocks are 4 (Izu) and 6 (Ebino) here. 
In either case, the
average number of foreshocks per mainshock
is very large. In the case of Izu, it is about 
256 within the range of 
30km and 20 days from the mainshock, which is an order of magnitude
larger than the number associated with swarm-unrelated mainshocks.

As can immediately be seen from Figs.\ref{HK:fig12}, the spatiotemporal pattern of 
these regions are quite different from the one in other regions shown in
Figs.\ref{HK:fig10}, 
which suggests that the property of swarm-related earthquakes 
might differ qualitatively from 
that of swarm-unrelated earthquakes. 
For example, foreshocks in the Izu swarm region
suddenly accelerate on the onset time of about a week, with the center
of activity about 5 km away from the mainshock position, the associated 
spatial correlation function  exhibiting a pronounced peak
at about $r\simeq 5$ km. This seismic pattern is not dissimilar to the
doughnut-like seismic pattern discussed by Mogi as
occurring preceding the mainshock  (Mogi doughnut) \cite{HK:Mogi}. However, since
the doughnut-like seismic pattern
as observed in Figs.\ref{HK:fig12} 
is not observed in our analysis of more generic
case of the wider region, at least in a statistically significant manner, 
it is likely to be related to the specialty of Izu (and Ebino)
region. Swarm activity is considered to be 
related to the activity of underground magma or water. Hence,
some of the statistical properties of swarm-related earthquakes 
might be better discriminated from 
those of more general swarm-unrelated earthquakes.

\begin{figure}[ht]
\begin{center}
\includegraphics[scale=0.65]{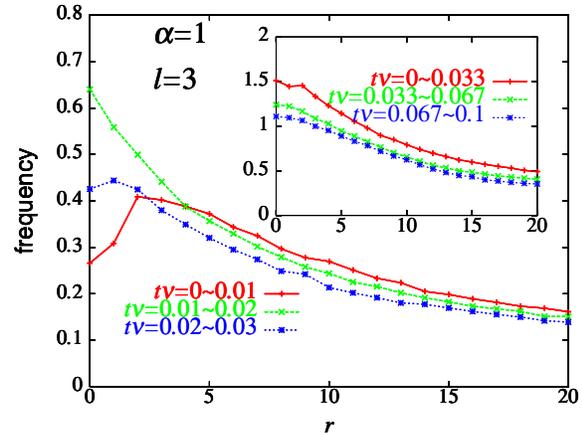}
\end{center}
\baselineskip 3.2mm
\caption{\small 
The time evolution of the spatial seismic distribution function before
the mainshock of $\mu >\mu_c=3$, calculated for the 1D BK model.
The inset represents a similar
plot with longer time intervals.
}
\label{HK:fig13}
\end{figure}

We examine next the corresponding spatiotemporal correlations of the
BK model. In the BK model, 
the distance from the epicenter is measured in units of blocks.
Fig.\ref{HK:fig13} 
represents the time evolution of the spatial distribution
of seismicity before the mainshock with $\mu >\mu_c =3$ 
for $\alpha=1$. Very much similar behaviors are observed
for other values of
$\alpha$. 
Similarly to the JUNEC data,
preceding the mainshock, 
there is a tendency
of the frequency of smaller events to be enhanced at and around 
the epicenter of the upcoming mainshock.
This was also observed previously by Shaw {\it et al\/} \cite{HK:Shaw92}.

Interestingly, however, as the mainshock becomes imminent, the frequency of
smaller events is suppressed 
in a close vicinity of the epicenter 
of the
upcoming mainshock, though it continues to be enhanced in the surroundings
(Mogi doughnut) \cite{HK:ScholzBook,HK:Mogi}.
We note that the quiescence observed 
here occurs only in a close vicinity of the 
epicenter of the mainshock, 
within one or two blocks from the epicenter, and only 
at a time close to the mainshock. 
The time scale for the appearance
of the doughnuts-like 
quiescence depends on the  $\alpha$-value:
The time scale of the onset of the doughnut-like 
quiescence tends to be longer for larger $\alpha$.
It is not clear at the present stage whether the doughnut-like
quiescence as observed here in the BK model has any relevance to the
one observed in Figs.\ref{HK:fig12} for certain earthquake swarms.

\begin{figure}[ht]
\begin{center}
\includegraphics[scale=0.65]{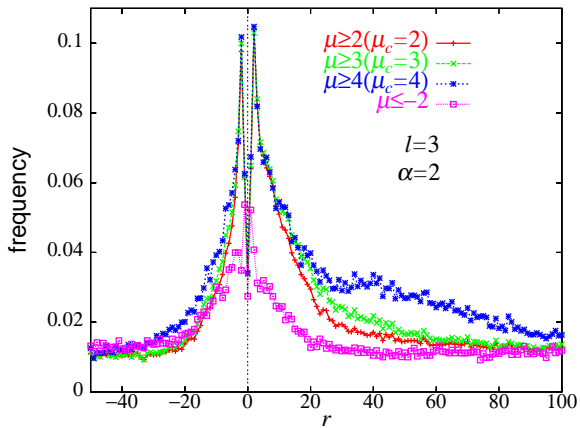}
\end{center}
\baselineskip 3.2mm
\caption{\small 
The frequency of seismic events preceding the events of various magnitude range
plotted versus $r$, 
the distance from the epicenter of the upcoming mainshock.
The curves correspond to the large events of $\mu >\mu_c=2,3$ and 4, and to
the smaller events of $\mu <-2$.
The direction in which the rupture propagates farther in the upcoming 
mainshock 
is always taken to be the positive direction $r>0$, whereas the direction 
in which the rupture propagates less
is taken to be the negative direction $r<0$.
The parameters are taken to be $\alpha=2$ and $l=3$, the time range being 
$0 \leq t \nu \leq 0.001$ before the mainshock.
}
\label{HK:fig14}
\end{figure}

As was discussed in detail in Ref.\cite{HK:MoriKawamura2}, in the
BK model,  the size of the
``hole'' of the doughnut-like quiescence  as well as its 
onset time scale have no correlation with the
magnitude of the upcoming event. 
In other words, 
the doughnut-like quiescence is not peculiar to
large events in the present model. 
This means that, by monitoring the onset of the ``hole'' in the
seismic pattern of the BK model, one can certainly deduce the time
and the position of the upcoming event, but unfortunately,
cannot tell about its magnitude. 
Yet, one might get some information about the magnitude of the upcoming event,
not from the size and the onset time of the ``hole'', but from the size
of the ``ring'' surrounding the ``hole''.
Thus, 
we show in Figs.\ref{HK:fig14} 
the spatial correlation functions before the mainshock
in the time range $0\leq t \nu \leq 0.001$ for the case of $\alpha=2$, 
with varying the magnitude range of the upcoming
event. In the figure, 
the direction in which the rupture propagates farther in the upcoming event
is always taken to be the positive direction $r>0$, whereas the direction in which the rupture propagates less
is taken to be the negative direction $r<0$. As can be seen from
the figure, although the size of the ``hole'' around the origin $r=0$ has
no correlation with the magnitude of the upcoming event as mentioned above,
the size of the region of the active seismicity surrounding this ``hole'' is
well correlated with the size and the direction of the rupture of the
upcoming event. This coincidence might enable 
one to deduce the position and the size
of the upcoming event by monitoring the pattern of foreshocks, although
it is still difficult to give a pinpoint prediction of the time of the upcoming
mainshock. We note that such a correlation between the size of the seismically active region
and the magnitude of the upcoming event was observed in the BK model 
in Ref.\cite{HK:Pepke}, and was examined in real earthquake data as well
\cite{HK:Kossobokov}.

\begin{figure}[bht]
\begin{center}
\includegraphics[scale=0.65]{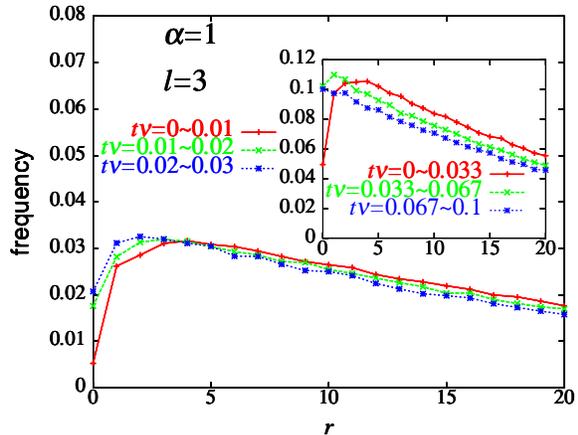}
\end{center}
\baselineskip 3.2mm
\caption{\small 
The time evolution of the spatial seismic distribution function before
the mainshock of $\mu >\mu_c=5$, calculated for the 2D BK model.
The inset represents a similar
plot with longer time intervals.
}
\label{HK:fig15}
\end{figure}

As shown in Fig.\ref{HK:fig15}, 
a very much similar behavior including the doughnut-like quiescence is observed
also in the 2D BK model. Note that the event frequency has been normalized
here 
by the measure factor $r$ associated with the area element of the polar
coordinate. The time scale of the onset of the doughnut-like quiescence
seems to be longer in 2D than in 1D.

\subsection{Spatiotemporal seismic correlations after the mainshock}

\begin{figure}[htb]
\begin{center}
\includegraphics[scale=0.65]{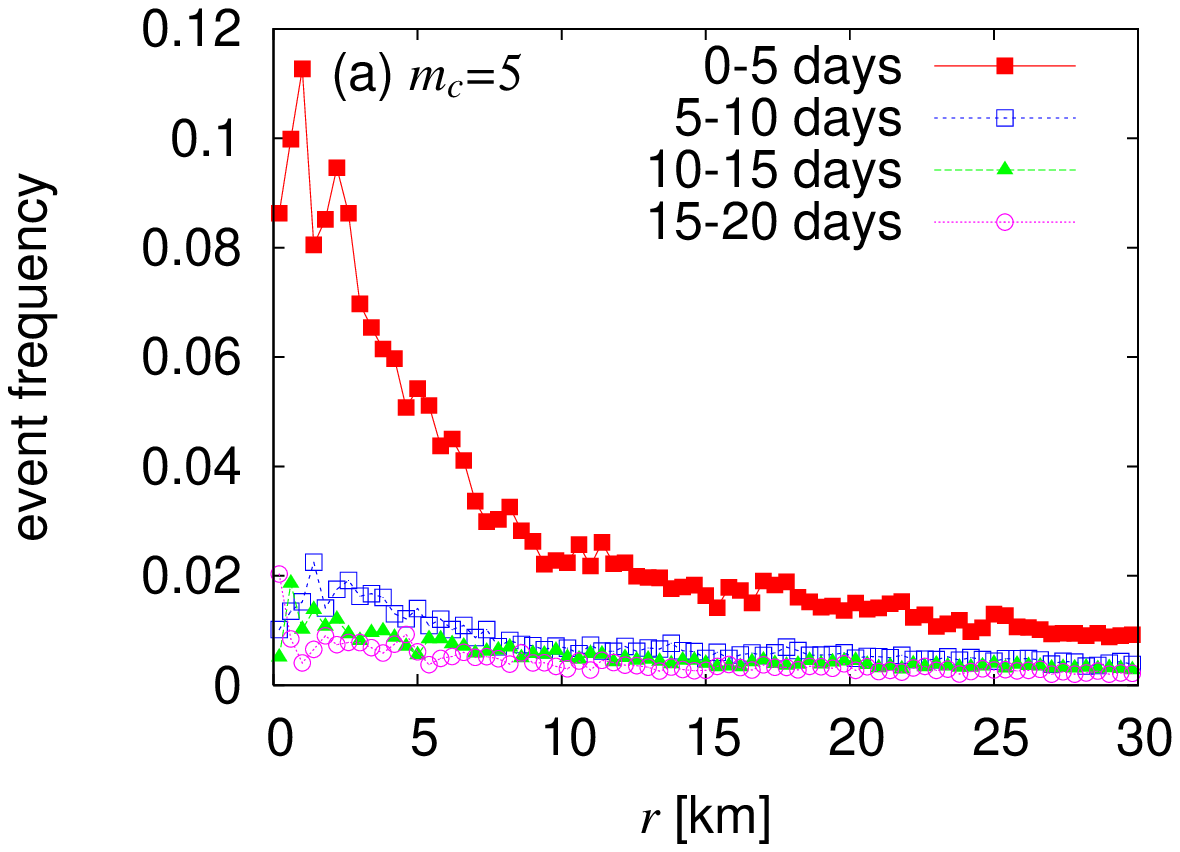}
\includegraphics[scale=0.65]{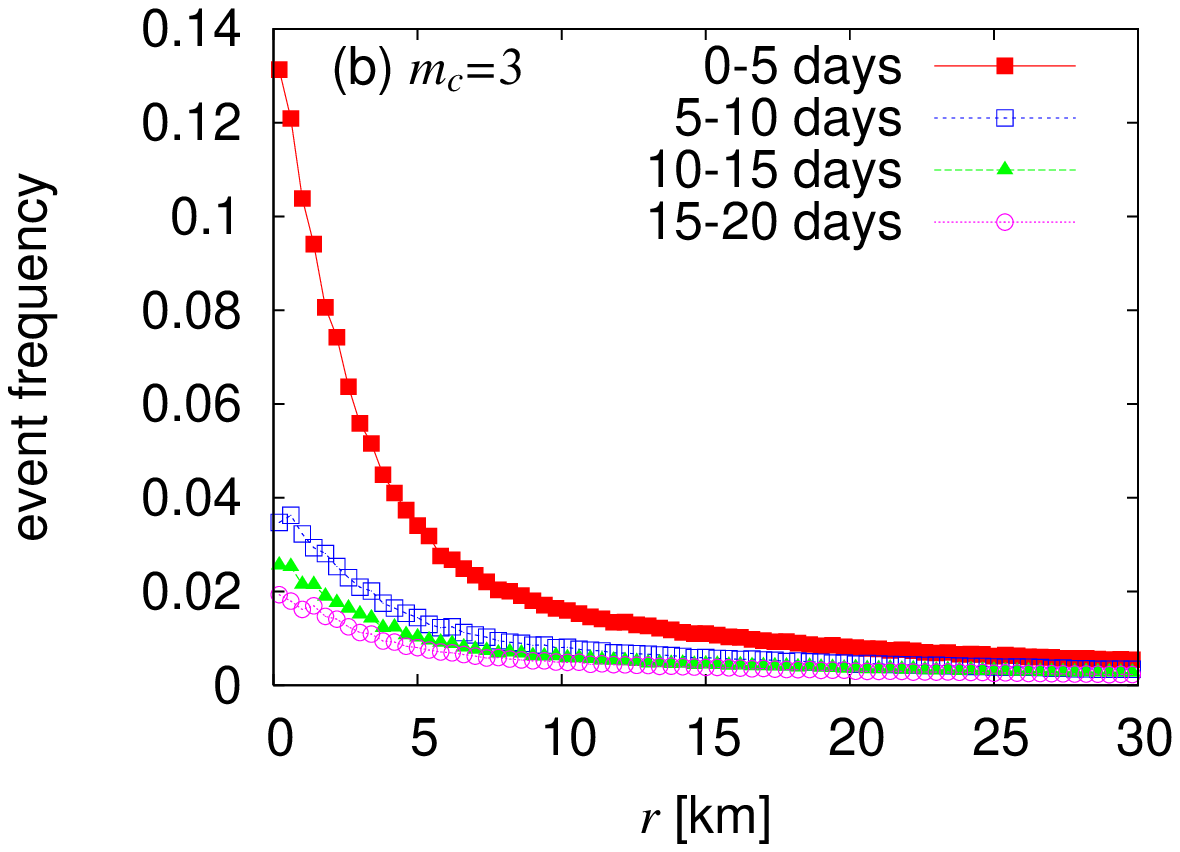}
\end{center}
\baselineskip 3.2mm
\caption{\small 
The time evolution of the event frequency of arbitrary magnitude after
the mainshock, 
plotted versus
the distance from the epicenter of the mainshock. The magnitude 
threshold for the mainshock is $m_c=5$ (a), and $m_c=3$ (b).
The data are generated from the JUNEC catalog.
}
\label{HK:fig16}
\end{figure}

In this subsection, we examine the spatiotemporal correlations
of earthquake events
{\it after\/} the mainshock.  
The time evolution of 
the spatial seismic pattern  calculated from the JUNEC catalog
are shown in Fig.\ref{HK:fig16}, with the magnitude threshold for the mainshock
$m_c=5$ (a), and $m_c=3$ (b).
As can be seen from these figures, aftershock activity is
clearly observed. The rate of the aftershock activity is highest 
just at the epicenter of the mainshock, not in the surroundings.
It is sometimes mentioned in the literature
that the aftershock activity is highest
near the edge of the rupture zone of the mainshock (see, {\it e.g.\/},
Ref.\cite{HK:ScholzBook}). However, this is not the case here.

\begin{figure}[htb]
\begin{center}
\includegraphics[scale=0.65]{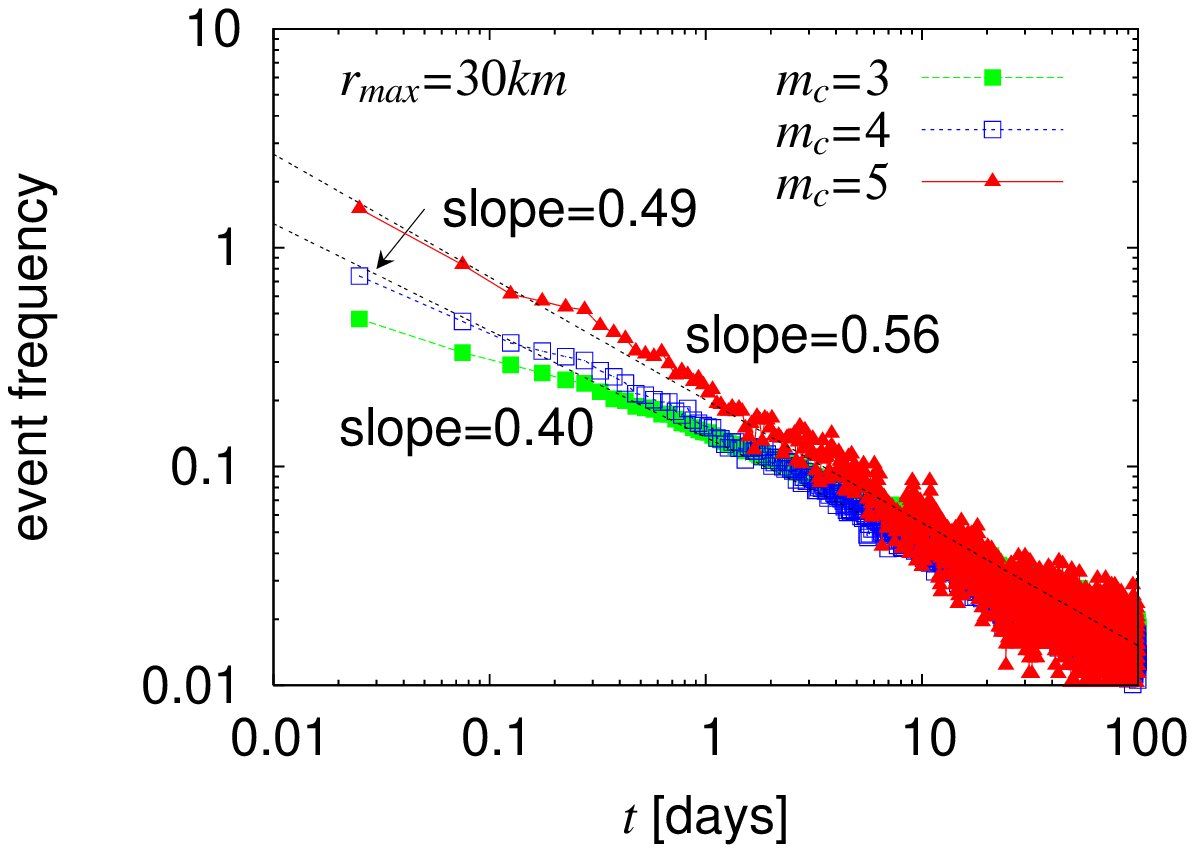}
\includegraphics[scale=0.65]{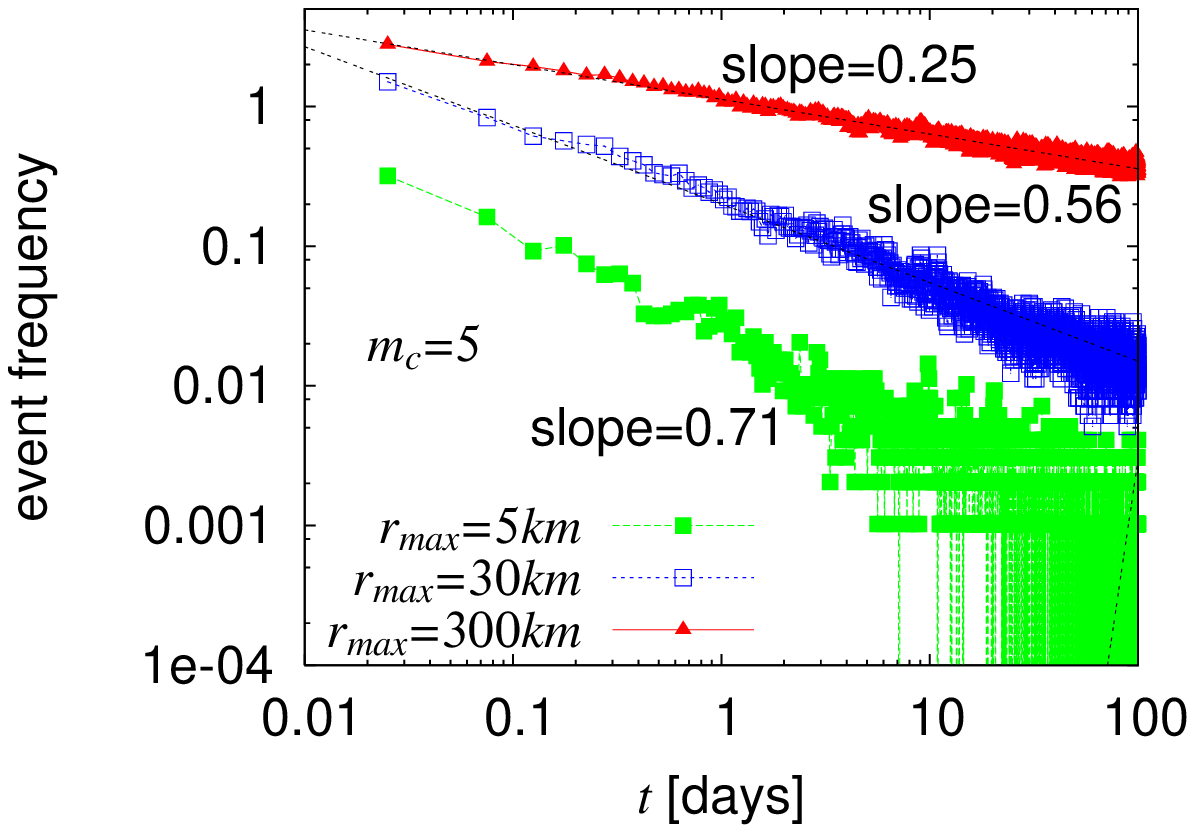}
\end{center}
\baselineskip 3.2mm
\caption{\small 
The frequency of seismic events of arbitrary magnitude after the mainshock, 
plotted versus the time after the mainshock on a double-logarithmic scale.
The data are generated from the JUNEC catalog.
In the upper figure, the magnitude threshold of the mainshock
is varied as $m_c=3,4$ and 5 with fixing the distance range of observation
$r_{max}=30$ km,
whereas, in the lower figure, the distance range of observation
is varied as $r_{max}=5,\ 30$ and 300 km
with fixing the magnitude threshold $m_c=5$.
}
\label{HK:fig17}
\end{figure}

In order to further clarify the relation with the  Omori law, we show
in Fig.\ref{HK:fig17} 
the time dependence of the frequency of aftershocks after 
the mainshock on a double-logarithmic scale: 
In the upper panel, the data with the distance range
$r_{max}=30$ km are shown with 
varying the magnitude threshold as $m_c=3,4$ and 5, while, in 
the lower panel, the data
with $m_c=5$ are shown  with varying the distance range 
as $r_{max}=5, 30$ and 300 km. In all cases analyzed, a straight-line behavior
corresponding to the power-law decay is observed (Omori law).
In contrast to the corresponding quantity before the mainshock,
the slope of the straight line, {\it i.e.\/}, the value of the Omori exponent, 
seems to depend on the distance range $r_{max}$ and the magnitude of the
mainshock $m_c$. 
As the distance range $r_{max}$ gets smaller and the magnitude
of the mainshock $m_c$ gets larger, the Omori exponent tends to get larger.
If one looks at the range $r_{max}=5$ km for mainshocks greater than 
$m_c=5$, the Omori exponent is about 0.71.

\begin{figure}[]
\begin{center}
\includegraphics[scale=0.65]{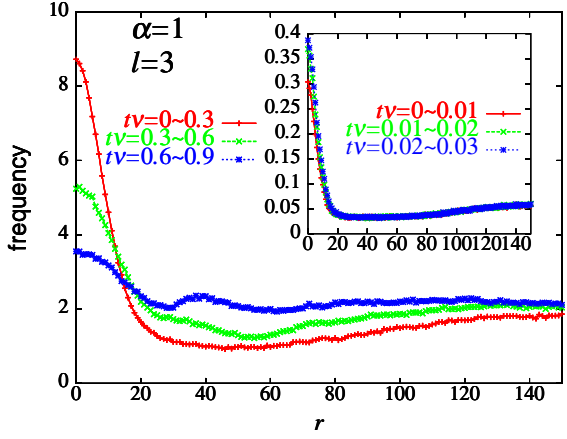}
\end{center}
\baselineskip 3.2mm
\caption{\small 
The time evolution of the spatial seismic distribution function after
the mainshock of $\mu >\mu_c=3$, calculated for the 1D BK model.
The inset represents a similar
plot with longer time intervals.
}
\label{HK:fig18}
\end{figure}

\begin{figure}[]
\begin{center}
\includegraphics[scale=0.65]{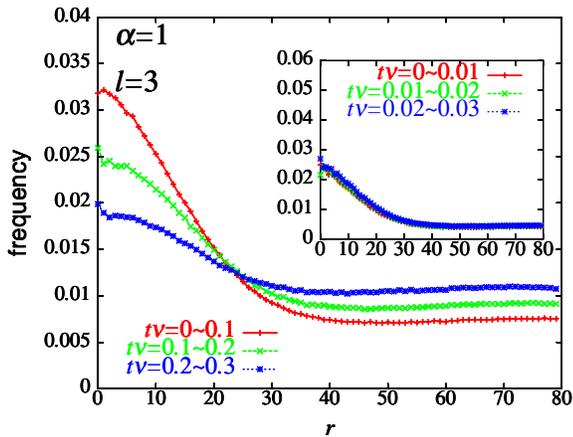}
\end{center}
\baselineskip 3.2mm
\caption{\small 
The time evolution of the spatial seismic distribution function after
the mainshock of $\mu >\mu_c=5$, calculated for the 2D BK model.
The inset represents a similar
plot with longer time intervals.
}
\label{HK:fig19}
\end{figure}

The corresponding spatiotemporal correlations
after the mainshock are calculated for the 1D BK model, and the results
are shown in Fig.\ref{HK:fig18} for the cases of $\alpha=1$.
As can be seen from the figure,
aftershock activity is clearly observed with the maximum rate occurring
at the epicenter of the mainshock. In contrast to the JUNEC data, the
$r$-dependence of the spatial seismic distribution is non-monotonic,
aftershock activity being suppressed in the surrounding region where
the rupture was largest in the mainshock.
The seismicity near the
epicenter is kept almost constant in time for some period after the mainshock, 
say, in the time range
$t\nu < 0.03$, which is in apparent contrast to the power-law decay as expected from the Omori law:
See the insets of Fig.\ref{HK:fig18}. At longer time scales, the seismicity 
near the epicenter seems to decay gradually, although
the decay observed here is not a power-law decay. 
A very much similar behavior is observed also for the
2D BK model: See Fig.\ref{HK:fig19}.
Hence, aftershocks obeying the Omori law is not realized in 
the BK model, as already reported \cite{HK:CL}. 
This is in apparent contrast to the observation for real faults.


Such an absence of 
aftershocks in the BK model might give a hint to the physical origin
of aftershocks obeying the Omori law, {\it e.g.\/}, 
they may be driven by the slow chemical process 
at the fault, 
or by the elastoplaciticity associated with the ascenosphere, 
{\it etc\/}, which are  not taken into account in the BK model.

\begin{figure}[]
\begin{center}
\includegraphics[scale=0.65]{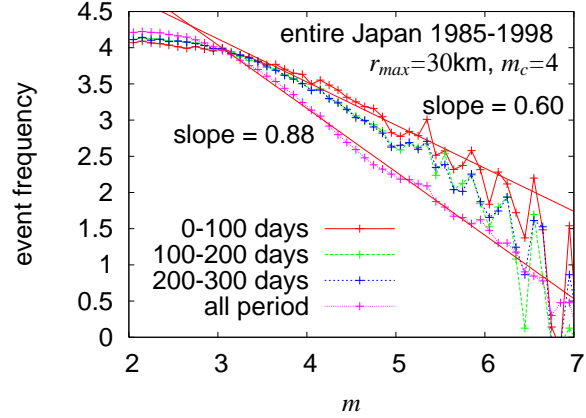}
\end{center}
\baselineskip 3.2mm
\caption{\small 
The time-resolved local magnitude distribution of earthquakes in Japan
before the mainshock of $m >m _c=4$, generated from the JUNEC
catalog. Preceding the mainshock, the $b$-value decreases considerably
from the long-time average value.
}
\label{HK:fig20}
\end{figure}

\subsection{The time-dependent magnitude distribution}

As an other signature of precursory phenomena,
we examine a ``time-resolved''  local magnitude distribution
for several time periods before the large
event.  Fig.\ref{HK:fig20} 
represents such a local magnitude distribution before large
earthquakes, calculated from the JUNEC catalog, {\it i.e.\/}, the
distribution of seismicity in the vicinity $r_{max}=30$ km of the
epicenter of the mainshock with its magnitude greater than $m_c=4$.
It can clearly be seen from the figure that the GR law
persists even just before the mainshock, whereas, 
preceding the mainshock,
the GR exponent $b$ gets smaller compared with the space- and time-averaged
$b$-value. For example, 100 days
before the mainshock, the $b$-value becomes about 0.60,
considerably smaller
than the averaged value $b\simeq 0.88$. Such a decrease of the
$b$-value was also reported in the literature \cite{HK:Suehiro,HK:Jaume}.

\begin{figure}[htb]
\begin{center}
\includegraphics[scale=0.65]{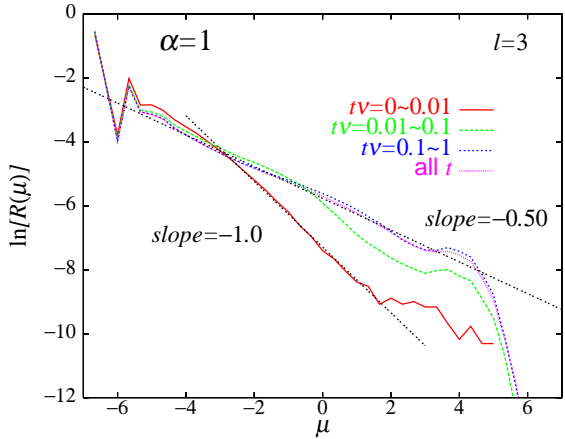}
\end{center}
\baselineskip 3.2mm
\caption{\small 
The time-resolved local magnitude distribution of the 1D BK model 
before the mainshock of $\mu >\mu _c=3$. 
}
\label{HK:fig21}
\end{figure}
\begin{figure}[htb]
\begin{center}
\includegraphics[scale=0.65]{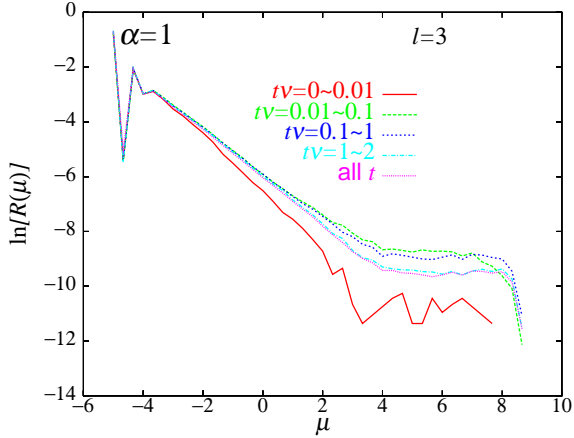}
\end{center}
\baselineskip 3.2mm
\caption{\small 
The time-resolved local magnitude distribution of the 2D BK model 
before the mainshock of $\mu >\mu _c=5$. 
}
\label{HK:fig22}
\end{figure}

For comparison, we show the corresponding local magnitude
distributions before the mainshock of the BK models both in 1D 
(Fig.\ref{HK:fig21})
and in 2D (Fig.\ref{HK:fig22}). 
The parameter $\alpha$ is taken to be $\alpha =1$.
In 1D,  only events with their epicenters within 30 blocks from the
epicenter of the upcoming mainshock are counted, while, in 2D, 
only events with their epicenters lying in a circle of its radius 5 blocks
centered at the epicenter of the upcoming mainshock are counted.
As can be seen from the figures, as the mainshock approaches, 
the form of the magnitude distribution changes significantly. 
In 1D, the apparent $B$-value describing the 
power-law regime tends to {\it increase\/}  as the mainshock
approaches, from the time-averaged value $B\simeq 0.50$ 
($b\simeq 0.75$) to the value
$B\simeq 1.0$ ($b\simeq 1.5$) just before the mainshock: It is almost doubled.
This tendency is opposite to what we have just found for 
the JUNEC catalog and several other observations for real faults 
\cite{HK:Suehiro,HK:Jaume}. However, a 
similar increase of the apparent $B$-value
preceding the mainshock was reported for some of real faults \cite{HK:Smith}. 
For the case of larger $\alpha >1$ (the data not shown here), 
the change of the $B$-value preceding the mainshock is still appreciable,  
though in a less pronounced manner.
More complicated behavior is observed in 2D. As the mainshock
approaches, the apparent $B$-value describing the 
power-law regime {\it slightly increases first, and then, 
increases\/} just before the mainshock.

\begin{figure}[htb]
\begin{center}
\includegraphics[scale=0.65]{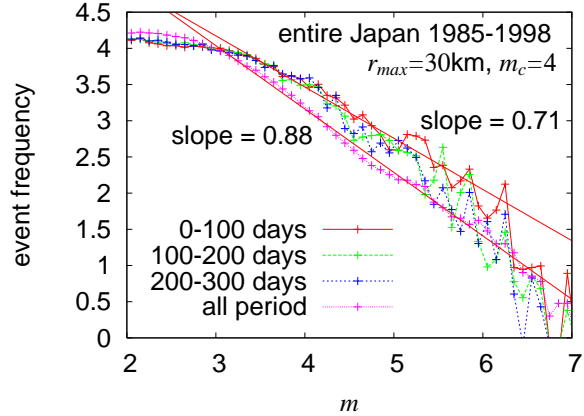}
\end{center}
\caption{\small 
The time-resolved local magnitude distribution of earthquakes in Japan
after the mainshock of $m >m_c=4$, generated from the JUNEC
catalog. 
}
\label{HK:fig23}
\end{figure}

Next, we analyze similar time-resolved  local magnitude distributions,
but {\it after\/} the large event.
Fig.\ref{HK:fig23} represents such a 
local magnitude distribution after the large
event calculated from the JUNEC catalog. In this case,
the deviation from the averaged distribution is relatively
small as compared with
the one observed before the mainshock, although there seems to
be a tendency that the observed
$b$-value decreases slightly after the mainshock.

\begin{figure}[htb]
\begin{center}
\includegraphics[scale=0.65]{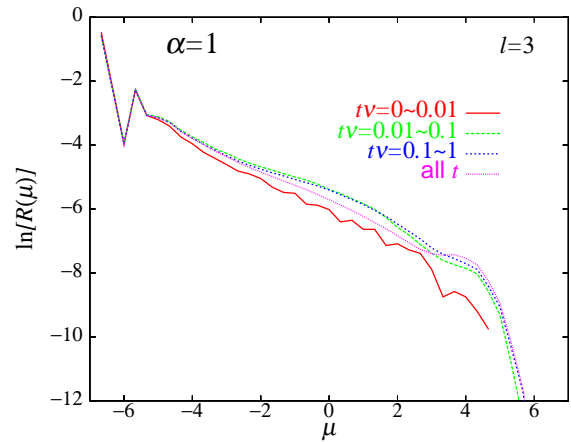}
\end{center}
\caption{\small 
The time-resolved local magnitude distribution of the 1D BK model 
after the mainshock of $\mu >\mu _c=3$. 
}
\label{HK:fig24}
\end{figure}

\begin{figure}[htb]
\begin{center}
\includegraphics[scale=0.65]{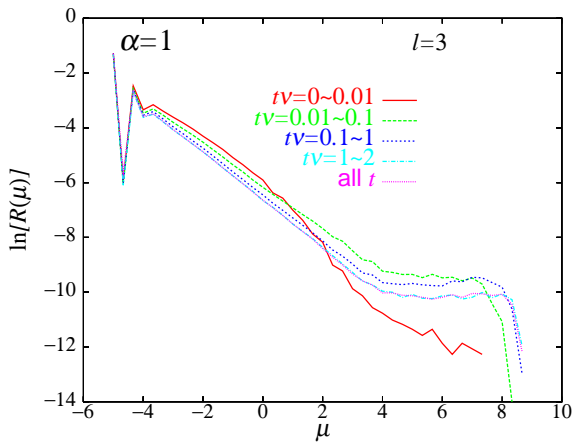}
\end{center}
\caption{\small 
The time-resolved local magnitude distribution of the 2D BK model 
after the mainshock of $\mu >\mu _c=5$. 
}
\label{HK:fig25}
\end{figure}

For comparison, we show the corresponding local magnitude
distributions after the mainshock for the BK models in 1D 
(Fig.\ref{HK:fig24})
and in 2D (Fig.\ref{HK:fig25}). 
The parameter $\alpha$ is taken to be $\alpha =1$.
As can be seen from the figures, 
the form of the magnitude distribution changes only little after the 
mainshock except that the weight of large events is decreased 
appreciably, particularly in 2D.

\noindent
\section{Summary and discussion}

In summary, 
we studied the spatiotemporal correlations of earthquakes both 
by the analysis of real earthquake catalog of Japan
and by numerical computer
simulations of the spring-block model in 1D and 2D. 
Particular attention was paid to the magnitude distribution, 
the recurrence-time distribution,
the time evolution of the spatial
distribution of seismicity before and after the mainshock, and
the time evolution of the magnitude distribution  before
and after the mainshock. 
Certain eminent features of the spatiotemporal correlations, including
foreshocks, aftershocks, 
swarms and doughnut-like seismic pattern, were discussed
in some detail. 

In our numerical simulations of the BK model, particular
attention was paid to the issue how the statistical properties of
earthquakes depend on the frictional properties 
of earthquake faults. 
We have found that when the extent 
of the velocity-weakening 
property is suppressed, the system tends to be more critical, while, as the
velocity-weakening property is enhanced, the system tends to be more 
off-critical with enhanced features of characteristic earthquakes.

Overall, the BK model tends to exhibit more characteristic or
off-critical statistical properties, particularly for large earthquakes,
than the real seismicity which exhibits much more critical
statistical properties. This discrepancy between the model
and the real seismicity has been recognized for some time now,
but its true cause has remained to be unclear. 

First, we need to
recognize that the earthquake catalog is taken not for a single
fault, but over many faults. There exists a suspicion that, even if
the property of a single individual fault is more or less characteristic 
the property obtained after averaging over many faults each of which
has different characteristics, becomes apparently characteristic as a
whole. If this is really the case, the real observation 
is not necessarily inconsistent with the observation
for the BK model, since the BK model deals with the property of
a single uniform fault. There is a claim that the extent of the criticality
of earthquakes might depend on the type of earthquake faults, {\it i.e.\/},
a matured fault with relatively regular fault zone
behaves more characteristic, while an inmatured fault with relatively 
irregular fault zone behaves more critical \cite{HK:Wesnousky}. 
Difficulty in testing such a hypothesis is that
the statistical accuracy of events for a single fault is rather limited, 
particularly for large events. We need to be careful because,
when the event number is not sufficient, an apparent
deviation from the criticality might well arise 
simply due to the insufficient
statistics, pretending a characteristic earthquake.

Other possibility is that, since smaller earthquakes are more or less
critical even in the BK model, in real seismicity, the critical behavior
might be limited to moderately large earthquakes which are contained 
in enough number in the
earthquake catalog, while very large earthquakes,
which are very few in number in the catalog, might be 
more or less characteristic. 
Anyway, the question of either critical or characteristic earthquakes
is one of major fundamental questions left in earthquake studies.

The BK model was found to exhibit several intriguing precursory phenomena 
associated with large events:
Preceding the mainshock, the frequency of smaller 
events is gradually
enhanced, whereas just before the mainshock it is 
suppressed only in a close vicinity of the epicenter of the upcoming mainshock
(the Mogi doughnut). The apparent $B$-value of the magnitude distribution 
increases significantly
preceding the mainshock.
On the other hand, the Omori law of aftershocks is not observed 
in the BK model. 

Some of these precursory phenomena observed in 
the BK model are also observed in
real earthquake catalog, but some of them are not.
For example, the enhancement of foreshock activity is observed in
common both in the BK model and in the JUNEC catalog.
By contrast, 
the doughnut-like quiescence generically observed in the BK model
is not observed in standard earthquakes
in the JUNEC catalog, although it is observed in 
certain earthquake swarms.

Here, in order to make a further link between the BK model and 
the real world, it might be of some interest to
estimate
various time and length scales involved in the BK model.
For this, we need to estimate  the units of time and length of the BK model
in terms of real-world earthquakes. Concerning the time unit 
$\omega ^{-1}$, we estimate it via the rise time of large earthquakes, 
$\sim \pi/\omega$, which is typically about 10 seconds. This gives an estimate 
of  $\omega^{-1} \sim 3$ sec.   Concerning the length unit $L$, 
we estimate it making use of the fact that the typical displacement 
in large events of our simulation is of order
one $L$ unit, which in real-world large earthquakes is typically 5 meters. 
Then, we get $L \sim 5$ meters. Since the loading rate $\nu'$ associated with 
the real plate motion is typically 5 cm/year, the dimensionless loading rate 
$\nu=N/(L\omega)$ is estimated to be $\nu \sim 10^{-9}$. If we remember that 
the typical mean recurrence time of large events in our simulation is 
about one unit of $\nu^{-1}$, the mean recurrence time of our simulation 
corresponds to 100 years in real world. 

In our simulation of the BK model, the doughnut-like quiescence was observed 
before the mainshock at the time scale of, say, 
$t\nu \lsim 10^{-2}$. This time scale corresponds to about 1 year. 
In our simulation, the doughnut-like quiescence was observed in the region only within a few blocks from the epicenter of the mainshock. To give 
the corresponding real-world estimate, we need the real-world estimate 
of our block size $a$.  In the BK model, 
the length scale $a$ is entirely independent of the length scale $L$, and 
has to be determined independently. We estimate $a$ via the typical 
velocity of the rupture propagation, $l a\omega $, which is about 3 km/sec 
in real earthquakes.
From this relation, we get $a \sim 3$ km.  
The length scale associated with the doughnut-like 
quiescence is then estimated to be 3$\sim $6 km in radius. 
If we remember that the rupture size of large events in our simulation with 
$\mu =\mu_c =3$ was about 60 blocks, the size of the rupture size of large 
earthquakes of our simulation corresponds to 180 km and more. 
This is comparable to the size of the rupture zone of real earthquakes
of their magnitude eight. Hence, the large events in the BK model
might correspond to exceptionally large earthquakes in real seismicity, which 
might explain the reason, at least partially, why the deviation from the
GR law observed in the BK model at larger magnitudes is hardly observed
in real seismicity. In real faults, the possible maximum 
size of the rupture zone might be limited by the fault geometry,
{\it i.e.\/}, by the boundary of a fault.

Of course, it is not a trivial matter 
at all how faithfully the statistical
properties as observed for the BK model represent those of
real earthquakes. We should be careful not to put too much meaning to the
quantitative estimates given above.

As an other precursory effect, 
the change of the apparent $B$-value is observed in the BK model,
{\it i.e.\/},
an increase in 1D or an initial decrease followed by a subsequent increase 
in 2D. In the JUNEC catalog, the $B$-value turns out to decrease 
prior to the mainshock consistently with several other observations for
real faults. Meanwhile, an increase of the $B$-value, which is 
similar to the one
observed in the 1D BK model, was reported in some real faults.
Thus, to elucidate the detailed mechanism 
behind the change of the $B$-value preceding the large event is an
interesting open question.

One thing seems to be clear: 
Much needs to be done before we understand the true nature of earthquakes.
We hope that 
further progress in statistical-physical approach to earthquakes,
combined with the ones in other types of approaches 
from various branches of science, 
would eventually promote our fuller understanding of earthquakes.

The author is thankful to Mr. T. Mori and Mr. A. Ohmura for their
collaboration and discussion. He is also thankful to Prof. B. Chakrabarti
for organizing the MOE conference and for giving me an opportunity of 
presenting a talk there and writing this article.

\end{document}